\documentclass{article}
\usepackage{arxiv}

\usepackage[utf8]{inputenc} 
\usepackage[T1]{fontenc}    
\usepackage{hyperref}       
\usepackage{url}            
\usepackage{booktabs}       
\usepackage{amsfonts}       
\usepackage{nicefrac}       
\usepackage{microtype}      
\usepackage{lipsum}		
\usepackage{graphicx}
\usepackage{cite}

\usepackage{array}
\usepackage{amssymb, amsmath, amsthm}
\usepackage{graphicx}
\usepackage{lmodern,url}
\usepackage{makecell} 
\usepackage{cancel}
\usepackage{multirow}
\usepackage{microtype}
\usepackage{lineno}
\usepackage{xspace}
\usepackage{xcolor}
\usepackage{siunitx}
\usepackage{todonotes}
\usepackage{booktabs}
\usepackage[ruled,vlined]{algorithm2e}
\usepackage{optidef}
\usepackage{mathdots}
\usepackage{subcaption}
\usepackage{csquotes}
\usepackage{caption}
\newcommand{\figref}[1]{Fig.~\ref{#1}}
\newcommand{\tabref}[1]{\tablename~\ref{#1}}
\graphicspath{{Figures/}}

\usepackage{amsmath}
\usepackage{tikz}
\usetikzlibrary{fit}
\usetikzlibrary{positioning}
\tikzset{%
  highlight/.style={rectangle,rounded corners,fill=red!15,draw,fill opacity=0.25,thick,inner sep=0pt}
}

\definecolor{mygreen}{RGB}{160, 242,182}

\newcommand{\RIC}[3]{\todo[inline,caption={},color=#1]{{\color{black}\textbf{#2}}\\ #3}}

\definecolor{mypurple}{RGB}{236, 223, 234}

\allowdisplaybreaks

\title{Mutational signatures and transmissibility of SARS-CoV-2 Gamma and Lambda variants}

\usepackage{authblk}

\author[1,*$\dagger$]{Karen Y. Oróstica}
\author[2,3,$\dagger$]{Sebastian Contreras}
\author[2,$\dagger$]{Sebastian B. Mohr}
\author[2,$\dagger$]{Jonas Dehning}
\author[2]{Simon Bauer}
\author[3,4]{David Medina-Ortiz}
\author[2]{Emil N. Iftekhar}
\author[1]{Karen Mujica}
\author[1]{Paulo C. Covarrubias}
\author[1]{Soledad Ulloa}
\author[1]{Andrés E. Castillo}
\author[5]{Ricardo A. Verdugo}
\author[1]{Jorge Fernández}
\author[3,4,$\ddagger$*]{Álvaro Olivera-Nappa}
\author[2,6,$\ddagger$*]{Viola Priesemann}

\affil[1]{Sub Department of Molecular Genetics, Institute of Public Health of Chile (ISP). Santiago, Chile.}
\affil[2]{Max Planck Institute for Dynamics and Self-Organization, Am Fa{\ss}berg 17, 37077 G\"ottingen, Germany.}
\affil[3]{Department of Chemical Engineering, Biotechnology and Materials, Universidad de Chile, Beauchef 851, 8370448 Santiago, Chile.}
\affil[4]{Centre for Biotechnology and Bioengineering, Universidad de Chile, Beauchef 851, 8370456 Santiago, Chile.}
\affil[5]{Human Genetics Program, Institute of Biomedical Sciences, Faculty of Medicine, University of Chile, Santiago, Chile}
\affil[6]{Institute for the Dynamics of Complex Systems, University of G\"ottingen, Friedrich-Hund-Platz 1, 37077 G\"ottingen, Germany.}
\affil[ ]{{$*$} Corresponding Authors: Karen Y. Oróstica (korostica@ispch.cl), Álvaro Olivera-Nappa (aolivera@ing.uchile.cl), and Viola Priesemann (viola.priesemann@ds.mpg.de)}
\affil[ ]{{$\dagger$} These authors contributed equally}
\affil[ ]{{$\ddagger$} These authors contributed equally}

\date{}

\renewcommand{\headeright}{ }
\renewcommand{\undertitle}{ }

\begin{document}
\maketitle

\begin{abstract}
The emergence of SARS-CoV-2 variants of concern endangers the long-term control of COVID-19, especially in countries with limited genomic surveillance. In this work, we explored genomic drivers of contagion in Chile. We sequenced 3443 SARS-CoV-2 genomes collected between January and July 2021, where the Gamma (P.1), Lambda (C.37), Alpha (B.1.1.7), B.1.1.348, and B.1.1 lineages were predominant. Using a Bayesian model tailored for limited genomic surveillance, we found that Lambda and Gamma variants' reproduction numbers were about 5\% and 16\% larger than Alpha's, respectively.
We observed an overabundance of mutations in the Spike gene, strongly correlated with the variant's transmissibility. Furthermore, the variants' mutational signatures featured a breakpoint concurrent with the beginning of vaccination (mostly CoronaVac, an inactivated virus vaccine), indicating an additional putative selective pressure. Thus, our work provides a reliable method for quantifying novel variants' transmissibility under subsampling (as newly-reported Delta, B.1.617.2) and highlights the importance of continuous genomic surveillance.
\end{abstract}

\section*{Introduction}

Despite widespread efforts on vaccination against COVID-19, the early lifting of restrictions and emerging variants of SARS-CoV-2 endanger a smooth transition from epidemicity to endemicity \cite{contreras2021risking,bauer2021relaxing,viana2021controlling,lavine2021immunological,cobey2021concerns}. Countries deploying only partially protecting vaccines and having limited resources for sustaining non-pharmaceutical interventions (NPIs) face additional challenges; larger fractions of the population remaining susceptible lead to higher levels of morbidity and mortality, and case under-reporting obscures the real extents of the spread \cite{contreras2020low,contreras2021challenges,bauer2021relaxing}. 

Higher COVID-19 incidence increases the risk of breeding new SARS-CoV-2 variants \cite{thompson2021incidence_and_escape}. 
Such variants could escape the partial immunity prevailing in parts of the population or have evolutionary advantages that facilitate their spread \cite{plante2021variant_gambit,van2020risk}. Genomic surveillance programs worldwide have reported more than 2.3 million SARS-CoV-2 genomes to the GISAID database \cite{GISAID}, where they are collected and shared. However, the capability to perform genomic surveillance effectively varies across countries, depending on the public policies and in low-to-middle income countries primarily on the resources to fund it \cite{cyranoski_alarming_2021,malick2021genomic,bartlow2021cooperative,helmy2016limited}. For example, in Chile, despite the governmental and private investments in genomic surveillance, the sequencing rate is around 400 samples per week. In these settings (hereafter referred to as subsampling), selecting which samples should be sequenced is fundamental for avoiding biases and misleading results. Thus, the presence of an entity coordinating the sampling and sequencing efforts (as the Chilean Public Health Institute) is mandatory.

The spread of COVID-19 in Chile has been remarkably heterogeneous, not only because of its geography and sparse urbanization but also because of the pronounced social inequalities \cite{mena2021science,gozzi2021estimating,bennett2021all,freire2021heterogeneous_spread_chile,contreras2020multigroup,castillo2020geographical}. The Chilean government has deployed an ambitious vaccination program \cite{shepherd2021vaccination_in_Chile,aguilera2021}, so that, to date, almost 60\% of the total population has been fully vaccinated \cite{owidcoronavirus}. However, despite its success in vaccination, the spread has not been completely controlled because of unsteady NPIs due to economic pressures \cite{asahi2021lockdowns_impacted_chile_economics}, reporting delays \cite{contreras2020statistically}, inefficient contact tracing \cite{MINSAL2021trazabilidad}, and the comparatively low protection against infection granted by the predominant vaccine. According to official sources, more than half of the administrated doses correspond to CoronaVac \cite{minsal_2021}, an inactivated SARS-CoV-2 vaccine with $65.9\%$ protection against infection \cite{jara2021effectiveness}. Furthermore, the partial isolation of certain regions of Chile and the fast connections to Santiago, the capital city, further favors the spread of locally generated variants \cite{gonzalez2021mutations_in_magallanes} or the insertion of new lineages in zones where there were no cases. The above highlights the importance of optimizing available genomic surveillance resources to timely alert policymakers about emerging threats, such as the Lambda lineage \cite{acevedo2021lambda_infectivity_chile,romero2021lambda_in_peru}. Recently, the Delta (B.1.617.2) variant has also been reported in Chile \cite{mora2021emergencia,vargas_2021}, and we are actively working in collecting sequencing data to incorporate it in revised versions of the manuscript.

Here, we quantified the contribution of different variants of SARS-CoV-2 to the spread of COVID-19 in Chile in 2021 and analyze the genetic drivers of the observable differences among lineages. We observed temporal variations of the genome (namely, total mutational load and individual mutations) in the samples collected, hinting at a selective pressure that prompts differentiation from the reference lineage. Growing post-infection and vaccine-induced immunity levels, and changing NPIs induce a varying susceptibility landscape, where certain variants might have comparative advantages. In that way, variants do not directly compete among each other but with the environment. Those that emerge manage to break the barrier imposed by the reduced susceptibility in the population, others either adapt or die out. Remarkably, our work includes the recently-emerged, but little-researched Lambda variant of interest. Quantifying the spreading potential of new variants through genomic surveillance enables the implementation of preventive measures. Thus, our work offers a framework for assessing the potential future impact of variants in the early stages of their spread using genomic data and Bayesian modeling.

\section*{Methods overview}

We sequenced whole SARS-CoV-2 genomes of samples from different Chilean regions using a MiSeq (Illumina) platform with a 300-cycle (total) reagent kit. We assessed sequencing quality with the FastQC program, v0.11.8, and used the IRMA (v0.9.3) and MAFFT (v7.458) software to respectively assemble and align the genomes \cite{shepard_viral_2016, katoh_mafft_2002}. To determine the lineage of each genome obtained, we used Pangolin v3.1.5 \cite{rambaut_dynamic_2020}.

To assess the relative transmissibility of the different variants in Chile, we proposed a Bayesian model, which simulates the spread of each variant separately using a discrete renewal process \cite{fraser_estimating_2007,Flaxman2020estimating, brauner_inferring_2020}. The disease spreads with an inferred time-dependent effective reproduction number $R_t$ \cite{dehning2020inferring}, with the addition that the reproduction number of each variant is modulated by a time-invariant factor $f_{\text{variant}}$. We set the Alpha variant as the reference, as it is well studied \cite{davies_estimated_2021}, defining its factor $f_{\rm Alpha}$ to be 1. The $f_{\text{variant}}$ variable, therefore, accounts for the relative transmissibility of the variants, i.e., their relative reproduction number compared to Alpha.
As data, we used the weekly relative share of the variants in all sequenced cases (i.e., the fraction a given variant represents of the total samples), assuming that these observations follow a multinomial distribution, and use the daily number of (largely non-sequenced) observed new cases to infer the absolute prevalence of the variants in time. Our model also included a small random influx of variants from abroad, which was essential to explain the sudden emergence of new variants among sequenced samples. Our method differs from the phylodynamic inference of population growth rates as implemented in BEAST 2 \cite{volz_bayesian_2018, bouckaert_beast_2019} in that it does not build phylogenetic trees, but only groups the different variants together, which simplifies significantly the inference. 
We herewith obtained an overall description of the spreading dynamics of the different variants over seven months.

In addition, we sought to understand the relationship between mutational patterns and transmissibility of the predominant variants, integrating sequencing data and variant-level spreading parameters. We analyzed the relationship between the accumulation of both specific and total mutations and the spread of the virus to detect patterns of co-occurrence of mutations over time.

Samples analyzed in this work were collected from hospitals belonging to the influenza surveillance network, strategically distributed across the country. All samples must have tested positive in an RT-PCR SARS-CoV-2 test with a Ct value lower than 25 and were sent to ISP in Santiago for sequencing under a strict cold transportation chain. Nevertheless, contingencies and other factors related to sample transportation may cause samples to be discarded, although samples were selected proportionally to regional COVID-19 incidence. Altogether, the above implies deviations from an ideal sampling (binomial distribution). Consequently, we incorporate a factor $\omega$ in our Bayesian framework, which penalizes non-ideal measures with more significant errors than expected under binomial sampling  (see Methods).

\section*{Results}

\subsection*{Quantification of the transmissibility of most predominant variants in Chile}

\begin{figure}[!h]
\hspace*{-1cm}
    \centering
    \includegraphics[width=6.5in]{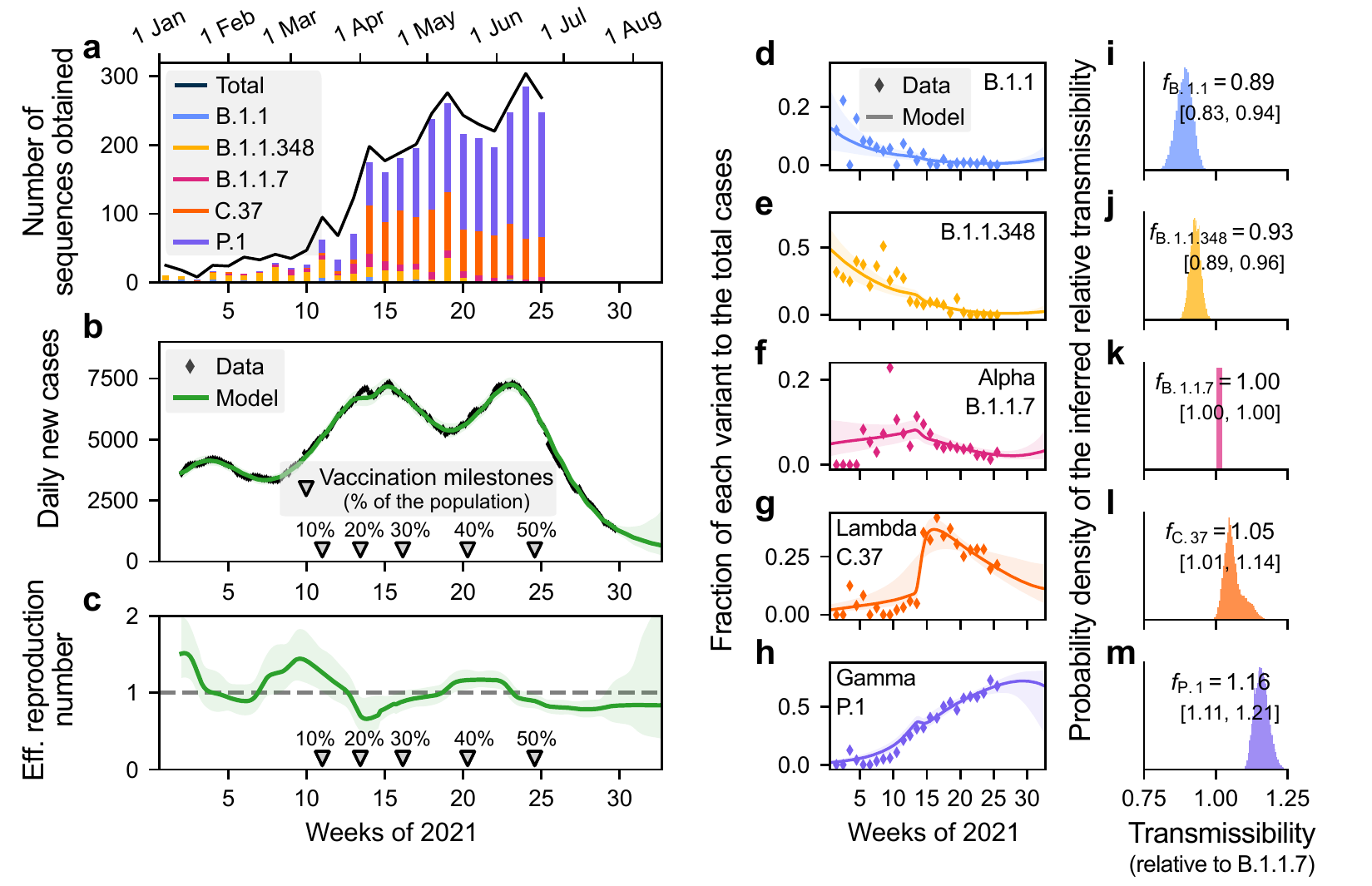}
    \caption{%
        \textbf{Bayesian inference enables individual assessment of the contribution of different SARS-CoV-2 variants to the spread of COVID-19. a.}
        Throughout 2021, five SARS-CoV-2 variants were identified as predominant in Chile, two considered Variants of Concern (VOC) by the WHO (Alpha, and Gamma), one Variant of Interest (Lambda), and two other unflagged lineages (B.1.1 and B.1.1.348). Assuming that the contribution of each variant to the spreading dynamics (\textbf{a--c}) is proportional to their share  (i.e., the fraction they represent of the total samples, \textbf{d--h}), we quantified their transmissibility compared to the Alpha variant (\textbf{i--m}). The Lambda and Gamma variants showed a 1.05 (95\% CI [1.01,1.14]) and 1.16 (95\% CI [1.11,1.21]) fold higher reproduction number than the Alpha variant. Other variants had a comparatively lower influence on the spread. Shaded areas in the \textbf{b--h} panels account for the 95\% credible intervals of the model fit. Complementary parameters and variables are summarized in Supplementary Figure~S1.
        }
    \label{fig:Figure_1}
\end{figure}

Since January 2021, we successfully sequenced 3443 SARS-CoV-2 genomes at the Chilean Public Health Institute (ISP), identifying 86 different lineages, of which only some have persisted over time. We filtered our dataset to analyze only those lineages representing at least 20\% of the total samples during one weekly observation period. Finally, we identified the Gamma (labeled as Variant of Concern, VOC), Lambda (labeled as Variant of Interest, VOI), Alpha (VOC), B.1.1.348, and B.1.1 lineages as predominant in the time frame analyzed (see~\figref{fig:Figure_1}a).

As of August 2021, the Gamma VOC, first reported in November 2020 in Manaus, Brazil \cite{faria_genomics_2021}, was the dominant variant in Chile, counting 1614 samples. It was followed by the Lambda VOI, with 790 samples identified from January. On the other hand, the Alpha (VOC, to date reported in 154 countries around the world \cite{Cov_lineages_B117}, has been detected only 122 times in Chile. In addition to those VOCs and VOI mentioned before, we have identified 253 samples classified as B.1.1.348 and 55 as B.1.1. The following sections will further characterize the epidemiological and genomic features of the circulating variants in Chile.

The Bayesian model fitted the daily number of cases well (\figref{fig:Figure_1}b) by adapting the effective reproduction number (\figref{fig:Figure_1}c) and also modeling the share of the different variants over time (\figref{fig:Figure_1}d--h). 
The emergence and sudden increase in the predominance of the Lambda variant around week 12 (cf. \figref{fig:Figure_1}g) is unlikely to be due solely to community transmission. As Lambda cases were zero or extremely low, this increase can be explained by an abrupt influx of cases (Supplementary Fig.~\ref{fig:Figure_S2}d), which acted as a seed for community transmission.

We found that the inferred relative reproduction number was the lowest for the non-VOC variants B.1.1 and B.1.1.348 (\figref{fig:Figure_1}i,j). From all the variants of concern and interest, our reference variant Alpha had the lowest reproduction number, followed by Lambda and Gamma with the highest reproduction number  (\figref{fig:Figure_1}k--m). In principle, knowing the base reproduction number of Alpha ($R_0 \approx 4.5$ \cite{davies_estimated_2021}) enables the estimation of other variants' base reproduction numbers by multiplying it by the corresponding factor $f$. 

\subsection*{Mutational load of the Spike gene correlates with variant transmissibility}

\begin{figure}[!h] 
\hspace*{-1cm}
    \centering
    \includegraphics[width=4.5in]{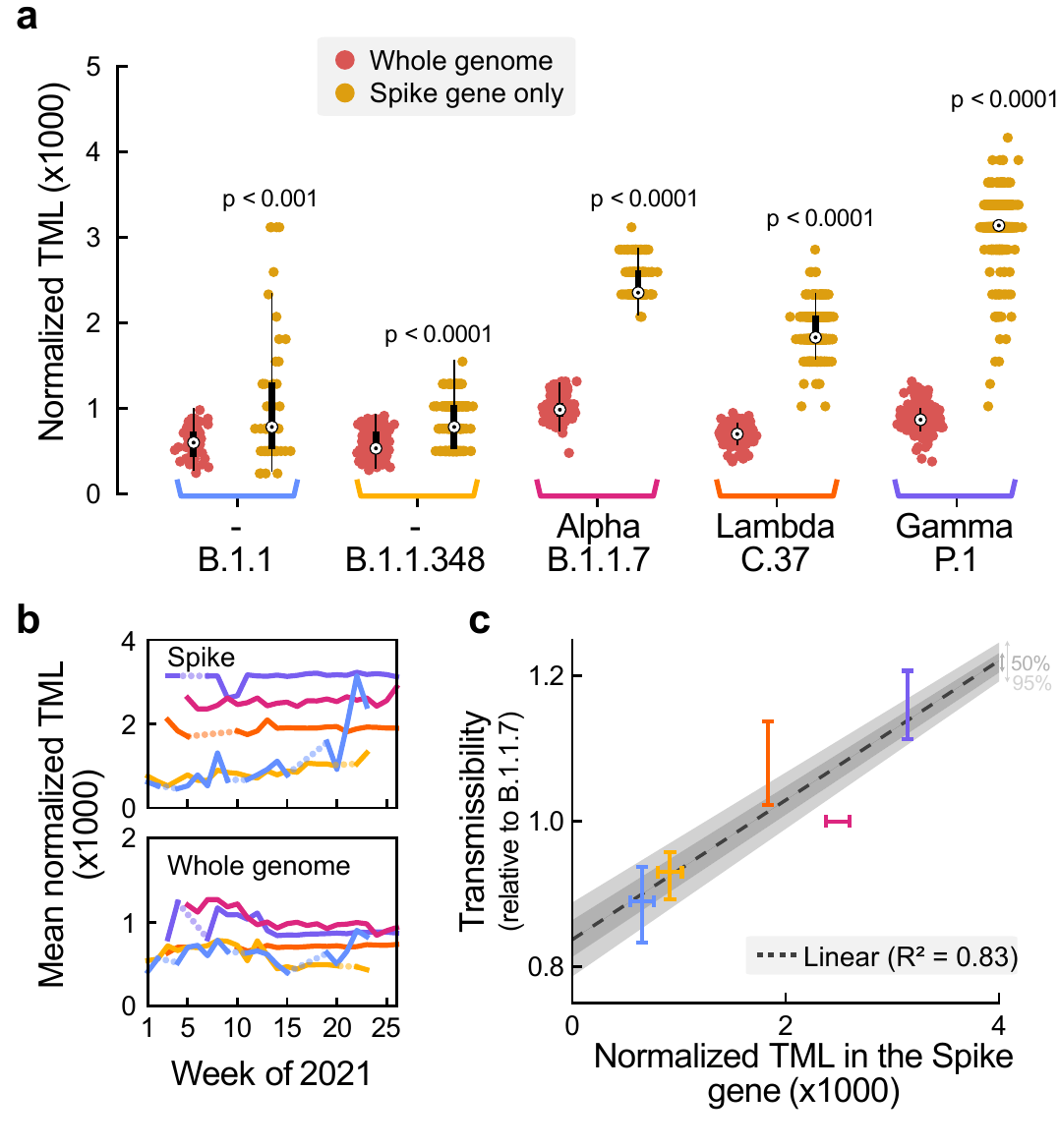}
    \caption{%
        \textbf{Predominant variants are enriched with mutations in the Spike gene. a.} When analyzing their normalized total mutational load (TML), namely, the number of mutations divided by the size of the gene x1000 (1~kbp), we observe that the number of mutations in the Spike gene (yellow) is above the average (red) for all variants herein analyzed. Furthermore, variants with higher normalized TML are consistent with those contributing more strongly to the transmission. The discreteness of the Spike gene mutational load is due to the shorter gene length. The white points denote the median, black boxes denote the interquartile ranges, and whiskers (thin black lines) extend until at most 1.5 times the length of the interquartile range. \textbf{b.} The most predominant variants do not show a considerable drift in their average TML over time. However, the TML could not account for the replacement of mutations, or other genetic dynamics, thus suggesting that it should not be used as a stand-alone measure of variability. Dotted lines account for weeks where the variants were not observed. \textbf{c.} The normalized TML in the Spike gene correlates positively with the relative contribution to the spread of the analyzed lineages. Furthermore, the discrepancy between observed TML and the linear regression is quite low, i.e., $R^2=0.83$. Errorbars denote 95\%. Vertical errorbars are those reported in Figure~1, and horizontal errorbars were estimated through bootstrapping. 
        }
    \label{fig:Figure_2}
\end{figure}

Next, we sought to understand the genomic drivers of the differences in spreading properties among variants through studying their mutations. These mutations could be insertions, deletions, or substitutions, and were typically missense, i.e., they caused an observable change in the generated amino acid sequence, thus likely having a functional effect in the translated protein \cite{teng_systemic_2020}. We then calculated the normalized Total Mutational Load (TML), i.e., the total number of mutations observed in the sequence compared to its reference, divided by the reference length. We calculated the normalized TML for both the whole genome and solely for the Spike gene, for the most predominant circulating variants in Chile (cf.~\figref{fig:Figure_2}a).

We observed a statistically significant enrichment in mutations in the Spike gene: Differences in the median value between the normalized TML for both the whole genome sequencing and Spike gene sequencing were significant for most predominant lineages in Chile (between $p\approx 0.001$ and $p\approx 0.0001$). Among the most predominant variants in Chile, Gamma had the highest number of mutations in the Spike gene, followed by Alpha, Lambda, B.1.1.348, and finally B.1.1 with the lowest TML (\figref{fig:Figure_2}a). The Spike gene showed a marked dispersion in the normalized TML in all samples compared to the whole genome. As the main differences between variants were on the degree of mutation enrichment in the Spike gene, we explored whether there is a correlation between the TML in the Spike gene and the relative transmissibility of the different variants.

The normalized TML in the Spike gene shows a marked linear correlation with the relative transmissibility of the most predominant lineages ($R^2= 0.83$,~\figref{fig:Figure_2}c). The Gamma variant had the highest total mutational load in the Spike gene compared to its whole genome (cf. \figref{fig:Figure_2}c). Furthermore, its relative reproduction number is markedly larger than other variants', and its share among collected samples has a marked increasing trend (cf.~\figref{fig:Figure_1}h). On the other hand, the Lambda variant was found to have a lower TML in the Spike gene than the Alpha variant while having a larger relative reproduction number. However, its relative prevalence in the population shows a decreasing trend. This might suggest that the spread of a certain variant would not only be related to the number of mutations but also to the composition of the accumulated mutation pattern, reflecting synergic or epistatic interactions between mutations.

\clearpage

\subsection*{Local samples differ significantly from reference genomes}
 
\begin{figure}[!h]
\hspace*{-1cm}
    \centering
    \includegraphics[width=6.5in]{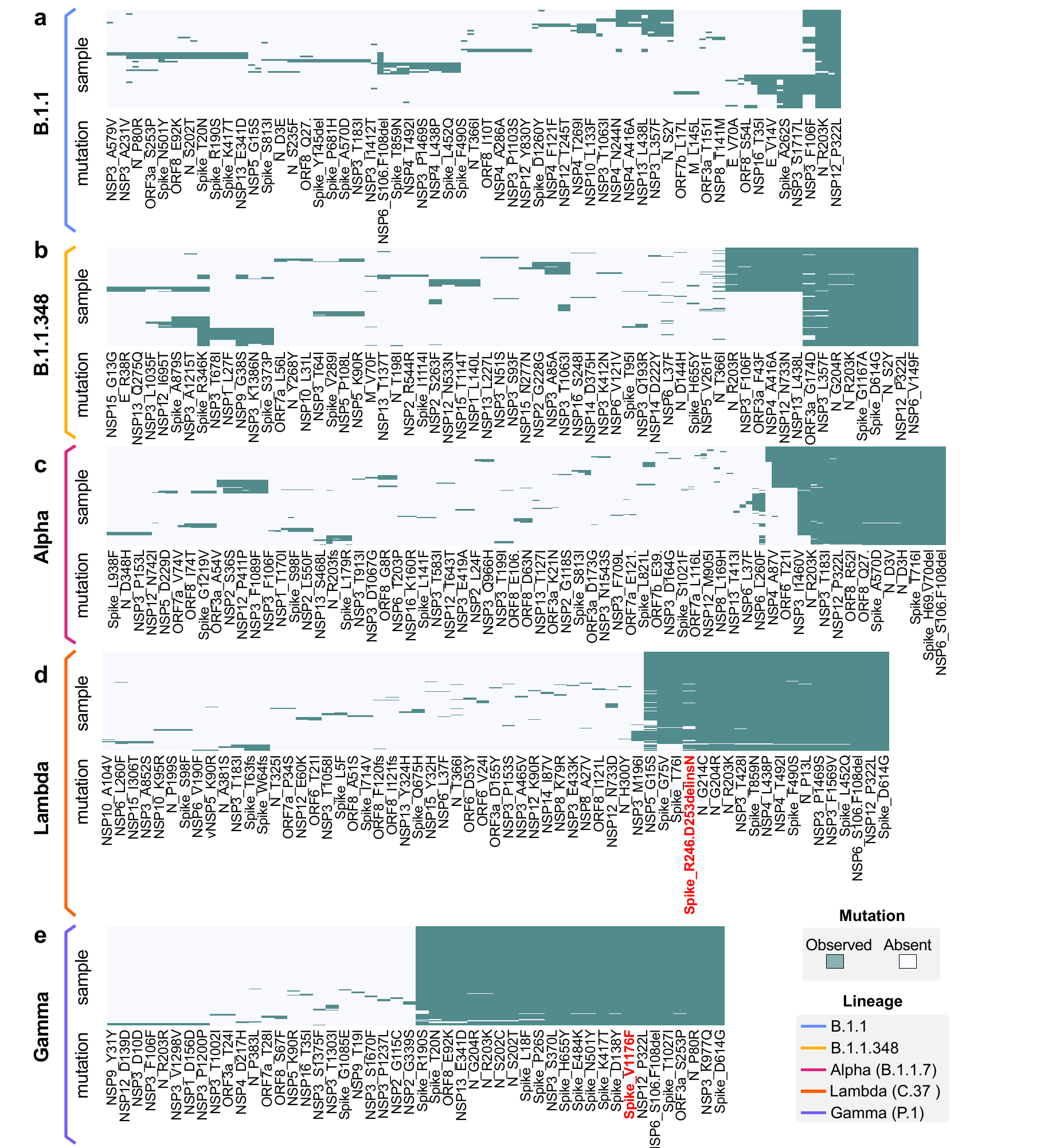}
    \caption{%
        \textbf{There exist different groups of co-occurring mutations under the same lineage category, indicating convergent evolution. a--e.} We analyzed the sequences of all samples collected for the predominant variants and analyzed whether they present a given mutation (filtered to be presented in at least 1\% of the samples). The mutations are grouped and sorted by co-occurrence, and observation frequency grows to the right of the plots.  Mutations highlighted in red have only been found in this study in the respective lineages. Supplementary Table~S1 summarizes the defining mutations for each lineage.
        }
    \label{fig:Figure_3} 
\end{figure}

Chilean samples of different variants systematically exhibit mutation patterns that are not present in the reference lineages, i.e., they drift from minimal list of defining mutations presented in Supplementary Table~S1. We selected the mutations present in at least 1\% of the samples of each lineage and sorted them according to their frequency of occurrence (\figref{fig:Figure_3}a--e). In the five most predominant lineages, besides the lineage-defining mutations we observe novel ones, not only restricted to the Spike gene. 

The B.1.1 lineage, first reported on March 1, 2020, has been detected in 47,500 sequences worldwide. However, in South America, Brazil alone has contributed with 20,702 sequences. Signature mutations of this lineage are in the N gene (R203K, G204R), the Spike gene (D614G), and in the ORF1b gene (P314L) \cite{Cov_lineages_B11}. Among the samples we have sequenced in Chile ($n=55$), the most predominant non-definitory mutations in the Spike protein were A262S and G1167A (\figref{fig:Figure_3}e and~\tabref{tab:Table_S1}).

The B.1.1.348 lineage is widely present in South America, and some countries in Europe \cite{GISAID}. Our analysis found four mutations in the Spike protein: D614G, G1167A, R346K, and S373P. The R346K mutation, together with A348T and N354K, has been suggested to improve the transmissibility of SARS-CoV-2 due to its higher binding affinity to the ACE2 receptor \cite{wang_characterizing_2020}. In addition, the S373P mutation in the RBD domain has been reported to escape immunity granted by mRNA vaccines partially and to decrease plasma therapy success \cite{mohammadi_novel_2021}. Therefore, both R346K and S373P mutations in B.1.1.348 are of particular interest and suggest the need to carefully observing their progression since they have been reported to favor the transmission of the virus and simultaneously reduce the effectiveness of vaccination \cite{mohammadi_novel_2021}.

The Alpha VOC, first detected in the UK in mid-2020, has among its defining non-synonymous mutations the  N501Y mutation in the RBD domain and deletion at positions 69 and 70 of the Spike protein, associated with enhanced transmissibility and pathogenicity \cite{volz_assessing_2021, frampton_genomic_2021}. In particular, its base reproduction number has been estimated to be around 4.5 \cite{davies_estimated_2021}. To date, we have analyzed 122 samples classified as the Alpha variant, and additionally, to its defining mutations, we found three new Spike mutations in about 20\% of the samples: G1219V, L938F, and S493P. The S493P is in close contact with the ACE2 binding region because it is located in the RBD domain. Furthermore, evolutionary analyzes have found that the S493P is under strong positive selection bias, altering human ACE2 binding affinity \cite{chakraborty_evolutionary_2021}.

The Lambda VOI has eight defining mutations in the ORF1a gene (T1246I, P2287S, F2387V, L3201P, T3255I, G3278S, P314L, and $\triangle$3675-3677) and seven mutations on the Spike gene ($\triangle$246-252, G75V, T76I, L452Q, F490S, D614G, and T859N) \cite{wink_first_2021, noauthor_outbreakinfo_nodate}. To date, 790 samples have been identified as Lambda from January 2021, making it the second most predominant variant in Chile. Additionally to its defining mutations, in Chilean Lambda samples, we found the R246\_D253delinsN mutation as a characteristic deletion and insertion of the lineage and no deletions in the ORF1a gene (\figref{fig:Figure_3}e). Other less predominant non-synonymous mutations are presented in~\figref{fig:Figure_3}e and summarized in Supplementary~\tabref{tab:Table_S1}.

The Gamma VOC has 21 defining mutations, of which ten occur on the Spike gene (L18F, T20N, P26S, D138Y, R190S, K417T, E484K, N501Y, D614G, H655Y and T1027I) \cite{naveca_covid-19_2021,faria_genomics_2021}, where the most relevant are K417T, E484K, and N501Y, located in the receptor-binding domain (RBD) of Spike \cite{faria_genomics_2021} (\tabref{tab:Table_S1}). Complementary to its defining mutations, we discovered among the Chilean Gamma samples an additional mutation in the Spike, V1176F, which has been reported to produce a more severe course of the disease \cite{nagy_different_2021}. Additionally, we found mutations on non-structural proteins such as ORF3a, NSP3, and NSP12, which play a relevant role in the evolution and spread of the virus.

\subsection*{Temporal drift of predominant variants suggests evolutionary pressure driven by vaccination}

As variants drifting from their lineage reference genome defined marked groups (cf.~\figref{fig:Figure_3}), we explored whether there was a temporal structure in their evolution. We analyzed the frequency of non-defining mutations observed among the samples for each lineage, selecting only those trends that present the most considerable variance.

\begin{figure}[!h]
\hspace*{-1cm}
    \centering
    \includegraphics[width=6.5in]{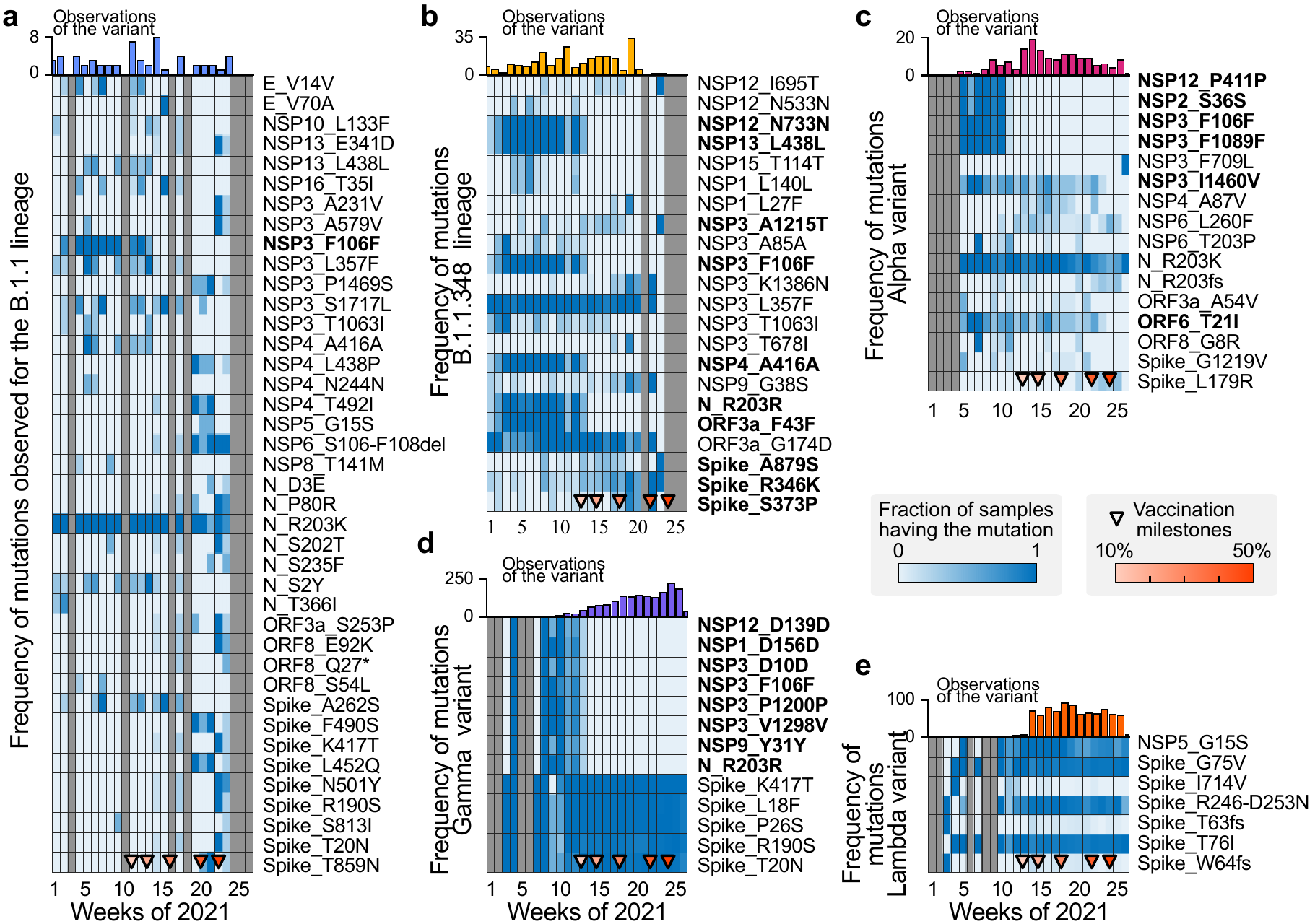}
    \caption{%
        \textbf{Signatures of the settlement, replacement, and selection of mutations in the different observed lineages of SARS-CoV-2.} Throughout 2021, the set of mutations that are present in the analyzed samples of the predominant lineages has changed. This temporal evolution of the mutational footprint of the lineages can be quantified by the proportion of the analyzed samples which present a given mutation. We selected mutations with the largest temporal variability for each lineage, and we present their evolution as a heat-map. \textbf{a--e:} Evolution of the fraction of the samples presenting a given mutation for the B.1.1 (\textbf{a}), B.1.1.348 (\textbf{b}), Alpha (\textbf{c}), Gamma (\textbf{d}), and Lambda (\textbf{e}) variants, respectively, with their number of observations. Triangle markers in the lower end of each heat-map account for the progress in vaccination. We can observe that, as vaccination progressed, some mutations were not frequent in the analyzed samples anymore. For example, the P411P (NSP12 gene), S36S (NSP2 gene), F106F and F1089F (NSP3 gene) mutations in Alpha (\textbf{c}) and the D156D (NSP1 gene), Y31Y (NSP9 gene), D139D (NSP12 gene), R203R (N gene) and D10D, P1200P, V1298V (NSP3 gene) in Gamma variants (\textbf{d}) were not frequently observed after reaching the $\approx 20\%$ fully vaccinated population milestone. On the other hand, mutations in the B.1.1 (\textbf{a}) variant are highly variable, which could be an subsampling artifact, and the mutational profile of the Lambda variant does not significantly change over time or with the progress of vaccination (\textbf{e}).
        }
    \label{fig:Figure_4}
\end{figure}

Samples belonging to B.1.1 and B.1.1.348 lineages (\figref{fig:Figure_4}a,b) presented the greater variability in their mutational profile. Although being highly dispersed, we did not observe a marked temporal variation in these trends (\figref{fig:Figure_2}b yellow and blue lines). Furthermore, their occurrence was less frequent than the other lineages (especially for B.1.1), suggesting that the subsampling-induced noise can explain parts of the variability. Nonetheless, these two lineages seemed to be strongly affected by vaccination, as they completely disappeared by the end of the analyzed period. As vaccination progressed, the NSP3\_F106F mutation for B.1.1 (~\figref{fig:Figure_4}a), and NSP12\_N733N and NSP13\_L438L mutations for B.1.1.348 (among other mutations highlighted in bold in~\figref{fig:Figure_4}b) steadily became less frequent. 

Lineages with the highest relative reproduction number and TML, i.e., Gamma, Lambda, and Alpha (cf. \figref{fig:Figure_2}), persisted throughout vaccine roll-out. For instance, several Alpha-lineage specific mutations became less frequent as vaccination progressed (\figref{fig:Figure_4}c): Mutations in the NSP2, NSP3, and NSP12 genes became less frequent after vaccination was noticeable, while mutations in the Spike gene started to appear by the end of the analyzed period. Similarly, Gamma variant lost several mutations in the NSP1, NSP3, NSP12, NSP9, and N genes, while keeping most of the mutations in the Spike domain (\figref{fig:Figure_4}d). Remarkably, the Gamma lineage steadily increased its share among the circulating variants, as more samples were identified as variants of it.

Finally, the circulating Lambda variants showed a relatively stable set of predominant mutations, as variability was only noticeable when Lambda samples were less frequent (\figref{fig:Figure_4}e, histogram). However, there was a slight tendency to incorporate more mutations in the Spike protein.
\clearpage
\section*{Discussion}

Using a Bayesian model, we estimated the relative transmissibility of Chile's most predominant SARS-CoV-2 variants and explored the genetic rationale behind the observed behaviors. Our results agree well with those reported in the literature and constitute ---to our knowledge--- the first approach to quantifying the spreading properties of the Lambda variant. We also update early estimates of the transmissibility of Gamma variant, incorporating novel mutations identified in our samples. Finally, we analyzed our results under the light of the genomic properties of the circulating lineages, correlating transmissibility with the number of mutations in the Spike gene and finding significant effects.

For inferring the transmissibility of the different variants, we made a set of assumptions aiming to overcome the challenges posed by subsampling in countries with limited genomic surveillance. Our methodology is general and easily adaptable to scenarios with more circulating variants and other countries. Even though more data would help to narrow the confidence intervals for our inferred parameters, the results were statistically consistent and agree well with the evidence already reported \cite{faria_genomics_2021, Campbell2021Increased}. For example, the mean increase in reproduction number for the Gamma variant compared to Alpha found in \cite{Campbell2021Increased} corresponds to $f_{\rm Gamma}=1.07$, which is close to our result  $f_{\rm Gamma}=1.16$. However, whether differences between variants are due to immune escape or enhanced transmissibility cannot be disentangled from the data available. 

The estimated reproduction number ($R_{\text{eff}}$) depends not only on the estimated spreading rate but also on the choice of the generation interval of infections (namely, the time between consecutive infections). However, whether changes in $R_{\text{eff}}$ are due to changes in the spreading rate or the generation interval cannot be disentangled from the data available. Therefore, if different variants should have different generation intervals, the estimations of their relative reproduction numbers $f_{\rm variant}$ would be affected. We performed our analysis during a period where $R_{\text{eff}}$ was close to one, minimizing the effect of potentially differing generation intervals between variants. However, differing generation intervals among variants would lead to larger differences on the estimated relative reproduction numbers $f_{\rm variant}$ in waves featuring a high $R_{\text{eff}}$. 

We assumed that the influx of infections (and thereby, of variants) was proportional to the COVID-19 incidence in neighboring countries and evenly distributed across all tracked variants. Currently, the influx corresponds to a tiny percentage of the total cases ($\leq\SI{5}{\percent}$, cf.~\figref{fig:Figure_1}b and \figref{fig:Figure_S2}a--e). However, more exact modeling would be required when neighboring countries have considerably more cases than the country of study, as the influx can considerably affect community spread.

Besides the modeling aspects, our results strongly rely on data and are thus affected by its quality. Samples analyzed in this work were collected from hospitals belonging to the influenza surveillance network, strategically distributed across the country, and proportional to the local COVID-19 incidence at the time of collection. Despite these efforts, logistic challenges related to sample transportation might have caused deviations from optimal representative sampling. However, as described in Methods, our model can mitigate these effects by penalizing highly correlated samples. Sequencing capabilities also increased considerably in the period analyzed as more resources were destined for genomic surveillance. Besides, as novel variants are officially defined a considerable time after their emergence, previous samples have to be reassigned (as was the case for the Lambda variant), posing a trade-off between analyzing newly sequenced samples or historical data. Overall, we can assume that the data is representative and that our modeling framework corrects any deviations from this trend.

Mutations in the observed samples were typically missense, i.e., caused changes at a protein level. Although all vaccines, and therefore vaccine-elicited antibodies, are targeted towards the SARS-CoV-2 Spike protein, mutational data suggests that evolutionary pressure was also exerted on other viral genes. This fact becomes evident after reaching the 20\% vaccination milestone, i.e., the eldest 20\% of the population was vaccinated, on weeks 13--14. Many earlier mutations in non-Spike genes disappeared during this transition, while others increased their frequency (cf. Supplementary~\tabref{tab:Table_S5} and Fig 4). However, we cannot infer a causal relationship between extinction and the appearance of mutations with the vaccination process with the data we have. 
Receding lineages (B.1.1 and B.1.1.348) tended to develop new Spike mutations before disappearing, while thriving lineages (Gamma and Lambda) tended to conserve and fix pre-existing Spike mutations. In contrast, all lineages, except for Lambda, consistently developed non-Spike mutations during vaccine roll-out. The remaining variants were probably selected through epistatic fitness of a restricted protein subgroup, particularly Spike (S) and the nucleocapsid (N) protein \cite{rochman_ongoing_2021}. On the other hand, our genomic data shows that both Gamma and Lambda lineages seem to have evolved successful Spike mutations previous to vaccination campaigns, suggesting a better survival of these variants when confronted to vaccine-elicited antibodies. 

Furthermore, our data suggests co-occurring mutations between the Spike and other viral proteins. This could also be putatively epistatic and driven by host adaptation. However, the evidence of extinction of non-synonymous mutations in non-Spike proteins suggests a selection mechanism that combines wiping-out of variants generated by genetic drift together with positive selection of fittest Spike and, possibly, nucleocapsid variants. This process seems to have eliminated virus variants that were unfit when confronted with vaccine-elicited antibodies and also carried inconsequential non-synonymous mutations, which resulted in the fixation and thriving of escape variants. However, whether there is causality behind this correlation should be separately studied. 

Specific mutations in the Spike gene of Gamma and Lambda variants were crucial for the survival of these variants during vaccine roll-out.
For the Gamma variant, Spike mutations have been associated to enhanced transmissibility (N501Y) and with partial immune escape (K417T and E484K) \cite{harvey_sars-cov-2_2021}.
For the Lambda variant, Spike mutations L452Q, F490S and deletion 246-252 conferred partial immune escape against neutralizing antibodies elicited by CoronaVac, and a higher infectiousness than the Gamma variant \cite{acevedo_infectivity_2021}. However, regarding its transmissibility, our results indicate it is not higher than that of the Gamma variant (cf. Fig~\ref{fig:Figure_1}l,m). 

Even though some of the lineages we report have been already studied in other countries and settings, Chilean samples differ from the GISAID/Pangolin reference genomes. In the context of transmissibility, variants showing a considerable drift from the original lineage can have enhanced properties. These properties make them more transmissible (i.e., $f_{\text{variant}}$ increases) or counteract the partial host population's natural immunity to the lineage, thus enabling its community transmission. The above suggests the necessity of thoroughly characterizing both genomic and epidemiological properties of variants and highlights the importance of performing genomic surveillance both in community infections and entry points to the country. In this sense, genomic surveillance may allow identifying hidden fast-spreading lineages before they become a threat. In turn, experts can propose timely preventive policies that are more accessible and entail fewer infection and death risks compared to later corrective policies. For instance, the use of vaccine-dependent green passports or mobility allowances could be instated or withheld following the advance or retreat of fast-spreading or vaccine-resistant lineages. Although variants that emerged abroad resulted from different selective pressures (host immune response, vaccination, NPIs), they thrived in the particular Chilean environment and thus must be controlled swiftly to avoid new pandemic waves \cite{castillo2020geographical}.

In summary, the methodology proposed in this work, supported by sufficient active genomic surveillance, can promptly detect all the circulating variants and estimate their transmissibility. Quantifying their contribution to contagion early on, we can assess whether they will endanger containment should they become predominant, and thus enable early eradication if they are evaluated to pose a threat. Therefore, through genomic surveillance, we could detect situations in which early control and lockdown could save us months of restrictions and fatalities.

\section*{Methods}

\subsection*{Nucleic acid extraction and amplification}

Nasopharyngeal samples, previously confirmed as positive for SARS-CoV-2, were used for total nucleic acid extraction using the automated system Zybio EXM 6000. Reverse transcription for cDNA synthesis was performed with SuperScript III One-Step RT-PCR System with Platinum Taq Kit, RNase OUT (Invitrogen) with 2 mM random primers and 4.5 $\mu$M DTT at 55$^{o}$C for 60 min. cDNA was amplified based on COVID-19 ARTIC Illumina Library Construction and Sequencing Protocol V.3 (Farr, 2020), generating two pools with 400 pb length amplicons covering the whole viral genome.

\subsection*{Library preparation}
DNA fragments from each pool were mixed together and library was prepared with Illumina DNA PREP kit (Illumina, San Diego, CA, USA), purified using Agencourt AMPure XP beads (Beckman Coulter, Brea, CA, USA) and quantified by Victor Nivo Fluorimeter (Perkin Elmer) using Quant-it dsDNA HS Assay Kit (Invitrogen). DNA libraries were sequenced in a MiSeq (Illumina) using a 300 cycles kit. Around 0.3 GB of data was obtained for each sample.

\subsection*{Whole Genome Sequence analysis}
Sequence quality was analyzed with FastQC software v0.11.8. Readings were filtered and trimmed with BBDuk software considering a minimum of 36 bases length and a quality above 20. Genome assembly was performed with IRMA software v0.9.3 using as a the reference sequence the NCBI entry NC\_045512.2. Genomes were aligned with MAFFT v7.458 and the lineages for the assembled sequences were assigned with Pangolin v3.1.5 \cite{rambaut_dynamic_2020}. Final genomes with epidemiological metadata were submitted to \url{https://www.gisaid.org/} for the final quality check and the corrected lineages.
We analyzed 3956 SARS-CoV-2 sequencing samples in the Chilean Public Health Institute (ISP) obtained from January 2021 to date, of which 3443 obtained good results in terms of quality and genome coverage. We used Pangolin to assign the variant classification for samples with good quality measures.

\subsection*{Determination of Total Mutational load} 

From the mutational data, we implemented an $m \times n$ mutation count matrix by considering all types of mutations and deletions. In the matrix, $m$ is the number of samples (2726, considering only those belonging to the five lineages studied herein), and $n$ is the number of genes (25 genes). Therefore, the value in entry $V_{i,j}$ indicates the number of mutations and deletions of gene $j$ in the sample $i$. Later, we computed the Total Mutational Load (TML), equivalent to the total number of mutations, divided by the length of the reference of the Spike gene and the whole genome, by 1~kb (kilobases) for each sample 
\begin{equation}
\text{TML}_{i}= \frac{1000}{w_i}  \cdot \sum_{j=0}^{j=m}V_{i,j} \label{eq:TML},
\end{equation}
where $w_i$ accounts for the sequence length, 3821 and 29903 Kbp for the Spike gene and whole genome, respectively. We then study whether there was a statistically significant enrichment of mutations in the Spike gene. For that, we first applied a Levene's test for evaluating whether, for a given lineage, the distributions of normalized TML for the whole genome and the Spike gene only have equal or different variances. Then, as the test confirmed that variances were different for all lineages, we used a non-parametric Mann–Whitney \textit{U} test to assess whether the medians of the categories were significantly different for every variant. Results for both assessments are summarized in Supplementary~\tabref{tab:Table_S2}.

\subsection*{Inferring the variant specific contribution to the spread} 

We built our model on top of our existing spreading dynamic model \cite{dehning2020inferring} to assess the relative transmissibility of the different variants in Chile. Given different data, this model can be easily adapted for other countries or time frames.

We simulated the spread of each variant independent whereby the susceptible pool $S$ was shared across the different variants. For each variant $v$ we computed the number of newly exposed $E_{v}$ iteratively given a prior distributions $E_{v,0}$ and the generation interval distribution $g$ with hyperprior $m$. This follows the work of \cite{fraser_estimating_2007,Flaxman2020estimating, brauner_inferring_2020}. To account for non pharmaceutical intervention or other measures against the spread we introduced the time-dependent effective reproduction number $R_t$, which is allowed a change every 14 days relative to the previous reproduction number. 

For each variant $v$ the effective reproduction number was modulated by the time-invariant factor $f_v$, called relative reproduction number in the text. Additionally to account for cases induced by travel we also add a small random influx $\Phi_v$ for each variant $v$ which was scaled by the reported case numbers in the neighboring countries $M_t$ (we used Argentina, Peru and Brazil). In discrete form the spreading dynamics in our model read as:

\begin{align}
    E_{v,t} &= \frac{S_t}{N} f_v R_{\text{base}, t} \sum_{\tau=0}^{10} E_{v,t-1-\tau} g_\tau + \Phi_{v,t} M_t,\\
    S_{t} &= S_{t-1} - \sum_{v} E_{v,t-1},\\
    g_\tau &= \text{LogNormal}(\tau;\mu=m,\sigma=0.4),\\
    m &\sim \text{Normal}(\mu=4,\sigma=1).
\end{align}

Whereby $N$ is the population size of our considered country (Chile). The susceptible pool gets initialized with the population size. The prior distributions for the initial new cases, influx and the time-invariant contribution factor were set to

\begin{align}
    E_{v,0} &\sim \text{HalfCauchy}(\sigma=100) \quad\forall v,\\
    f_v &\sim \text{LogNormal}(\mu = 1, \sigma = 1) \quad \text{for } v \in \{ \text{B.1.1, B.1.1.348, Gamma, Lambda} \},\\
    f_{\text{Alpha}} &= 1,\\
    f_{\text{others}} &= f_{\text{others}}(t) \\
    \Phi_{v,t_w} &\sim \text{HalfStudentT}_{\nu=4}(\sigma=0.0005) \quad\forall v, \forall t_w.
\end{align}

The external input $\Phi_{v,t_w}$ was modeled in a weekly fashion, indexed by $t_w$, to decrease the number of variables to be estimated. In addition to the five variants mentioned in the main text, we also include in our model the share of sequenced cases not categorized into these five variants ($f_{\text{others}}$). In contrary to the other five main variants, the relative reproduction number of these other variants is allowed to vary over time (described later).

Let $y_{v,t}$ be the measured number of samples successfully sequenced (from samples having a positive PCR test), corresponding to variant $v$. Let $n_t$ be the total number of sequenced samples and $\tau_{v,t}$ the inferred relative case numbers of the variant $v$ at time $t$ compared to the total non variant case numbers. If we model the number of samples $y_{v,t}$ corresponding to a variant $v$ as a multinomial random variable, and assuming that samples collected for sequencing are independent, we can build the multinomial likelihood function for our model with our real world data $y$ and $n$ and the fraction of variant $\tau$ from the model:

\begin{align}
    y_{v,t} &\sim \text{Multinomial}(p_v = \tau_{v,t}, n=n_t) \quad \forall t.
\end{align}

The fraction $\tau_{l,t}$ is obtained from the model by the fraction between daily cases of a variant $v$ and total daily cases.

\begin{align}
    \tau_{v,t} = \frac{E_{v,t}}{\sum_{v} E_{v,t}}
\end{align}

However in our model we do not use this multinomial likelihood function but instead parameterize our model using the conjugate distribution, the Dirichlet distribution. In theory it is equivalent to using the multinomial distribution. The advantage is that we can add a factor $w$ that parameterizes an eventual non-optimal sampling strategy, for example, samples that are not being perfectly randomized across the country but are correlated to some extent. This has mathematically the consequence that the measured fractions $y_{n,t}/n_t$ are all reduced by a factor $w$. Thus, the resulting likelihood function is given by:

\begin{align}
    \tau_{v,t} &\sim \text{Dirichlet}(\alpha=w \cdot  \frac{y_{v,t}}{n_t}  + 1) \quad\text{with}\\
    w &\sim \text{Gamma}(\alpha=5,\beta=5)\\ 
\end{align}

To infer the slowly changing reproduction number we introduce sigmoidal change points relative to the previous reproduction number whereby the priors for the date of occurrence $d$ of the change point $c$ are set every 14 days. The transient length $l$ such as the date $d$ of each change point $c$ are defined relatively flat to express our uncertainty in these values.

\begin{align}
    R_{\text{base}, t} &= \exp \left(\sum_c \gamma_{c}(t)\right)\\
    \gamma_{c}(t) &= \frac{\Gamma_c}{1+e^{-4/l_{c} \cdot (t - d_{c})}}\\
    d_c &\sim \text{Normal}(\mu=14, \sigma=5) &\forall c\\
    l_c & \sim \text{Normal}(\mu=20,\sigma=6) &\forall c\\
    \Gamma_c &\sim \text{Normal}(\mu=0,\sigma=0.2) + \Gamma_{c-1}  &\forall c\neq0\\
    \Gamma_0 &\sim \text{Normal}(\mu=1,\sigma=0.2)
\end{align}

For the five variants that we focused on in the main text, $R_{\text{base}, t}$ is multiplied by an time-invariant relative reproduction number $f_v$. For the spread of the `other variants' that we modeled separately, we multiplied this $R_{\text{base}, t}$ by a time dependent $f_{\text{others}}(t)$ as the mixture of variants can slowly change over time. We assumed the this change is slower than the $R_{\text{base}, t}$:

\begin{align}
    f_{\text{others}}(t) &= \exp \left(\sum_c \gamma_{\text{others},c}(t)\right)\\
    \gamma_{\text{others},c}(t) &= \frac{\Gamma_{\text{others}, c}}{1+e^{-4/l_{\text{others},c} \cdot (t - d_{\text{others},c})}}\\
        d_{\text{others},c} &\sim \text{Normal}(\mu=14, \sigma=5) &\forall c\\
    l_{\text{others},c} & \sim \text{Normal}(\mu=20,\sigma=6) &\forall c\\
    \Gamma_{\text{others},c} &\sim \text{Normal}(\mu=0,\sigma=0.2) + \Gamma_{\text{others},c-1}  &\forall c\neq0\\
    \Gamma_{\text{others},0} &\sim \text{Normal}(\mu=1,\sigma=0.2)
\end{align}

Additional to the sequenced samples we constrain our model using the publicly reported case numbers (in Chile) $C_t$ aggregated by the Johns Hopkins University~\cite{JHU}. We sum over the newly infected pools for all variants to obtain the total number of new infections $E_t = \sum_v E_{v,t}$. These are than delayed with the LogNormal kernel with mean delay $D$ to account for a reporting delay and further modulated by a weekly absolute sinus function parameterized by an amplitude $h_w$ and an offset $\chi_w$. 

\begin{align}
   \hat{C_t} & = (1-h(t)) \cdot  \sum_{\tau=1}^{T} E_{t-\tau} \cdot \text{LogNormal}(\tau; \mu = D, \sigma = 0.3) \\
   &= (1-h(t)) \cdot \sum_{\tau=1}^{T} E_{t-\tau} \frac{1}{0.3\cdot\tau\sqrt{2\pi}}e^{-\frac{\left(\log(\tau)-\log(D)\right)^2}{2\cdot0.3^2}} \\
   D &\sim \text{LogNormal}(\mu = 10, \sigma = 0.2)
   \quad\text{ and with}\\
    h(t) & = (1-h_w) \cdot \left(1 - \left|\sin\left(\frac{\pi}{7} t- \frac{1}{2}\chi_w\right)\right| \right)
\end{align}

The likelihood given the reported case numbers $C_t$ is than modeled by a StudentT distribution and quantifies the similarity between model outcome and the available real-world time series. The scale factor $\kappa$ heuristically incorporates the measurement noise.

\begin{align}
    C_t &\sim \text{StudentT}_{\nu = 4}( 
        \mu= \hat{C_t},
        \sigma = \kappa \sqrt{\hat{C_t}}) \quad \text{with} \\
        \kappa &\sim \text{HalfCauchy}(\sigma=10)
\end{align}

For a complete list of model parameters and priors see \tabref{tab:parameters} and \tabref{tab:priors} respectively.

\section*{Author Contributions}

Conceptualization: KYO, SC, SBM, JD, SB, ÁO-N, VP\\
Methodology: KYO, SBM, JD, VP\\
Software: KYO, SBM, JD\\
Validation: KYO, SC, JD, SBM, SB, ÁO-N, JF, VP\\
Formal analysis: KYO, SC, JD, SBM, DM-O\\
Investigation: KYO, SC, JD, SBM, AC, KM, PC, SU, AC, DM-O\\	
Resources: KYO, JF, AC, KM, SO, PC\\
Data curation: KYO, JF, AC\\
Writing - Original Draft: KYO, SC, SB, SBM, JD, ENI, ÁO-N, AC, KM, PC, SU, RAV\\	
Writing - Review \& Editing: KYO, SC, JD, SBM, ENI, JF, VP, RAV\\
Visualization: KYO, SC, JD, SBM\\	
Supervision: ÁO-N, JF, VP \\	
Project administration: KYO, SC, ÁO-N, VP\\
Funding acquisition: ÁO-N, JF, VP.

\section*{Data availability}
Some source code for data generation and analysis is available online on GitHub \url{https://github.com/Priesemann-Group/covid19_inference}.
Sequencing of the test was done at the Chilean Public Health Institute (ISP). All genomes sequenced by ISP are hosted in the GISAID Initiative \cite{GISAID}. Additionally for the Bayesian inference we used the daily case reports for Chile, Brazil, Argentina and Peru aggregated by the Johns Hopkins University \cite{JHU}.
\RIC{mypurple}{Note}{If you already want to have a look into the code for the Bayesian analysis feel free to write a message to SBM at sebastian.mohr@ds.mpg.de. It will be publicly available at a later point in time.}

\section*{Acknowledgments} 
We thank the Priesemann group for exciting discussions and for their valuable input, and the Molecular Genetics and Viral Diseases Sub Departments of the ISP for their valuable assistance. We thank Anamaria Sanchez Daza for carefully reading, commenting, and improving the manuscript. All authors with affiliation (2) received support from the Max-Planck-Society. SC, DM-O, and \'AO-N received funding from PIA-FB0001, ANID, Chile. JD and SBM received funding from the "Netzwerk Universitätsmedizin" (NUM) project egePan (01KX2021). This project is also supported by grant no. COVID0557 by ANID.

\newpage
\renewcommand{\thefigure}{S\arabic{figure}}
\renewcommand{\figurename}{Supplementary~Figure}
\setcounter{figure}{0}
\renewcommand{\thetable}{S\arabic{table}}

\setcounter{table}{0}
\renewcommand{\theequation}{\arabic{equation}}
\setcounter{equation}{0}
\renewcommand{\thesection}{S\arabic{section}}
\setcounter{section}{0}
\section*{Supplementary Information}

\begin{figure}[!h]
    \centering
    \includegraphics[width=3.25in]{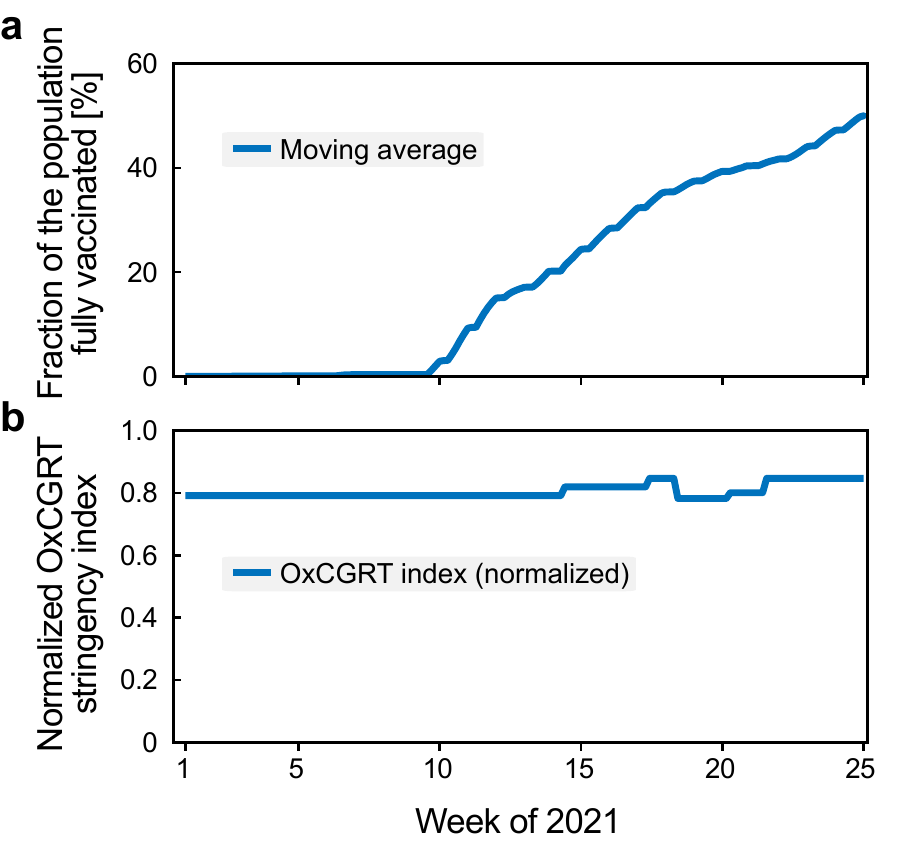}
    \caption{%
        \textbf{Progress of the vaccination program in Chile and the OxRCTT stringency index during vaccine rollout.}
        }
    \label{fig:Figure_S1}
\end{figure}

\begin{figure}[!b]
    \centering
    \includegraphics[width=\linewidth]{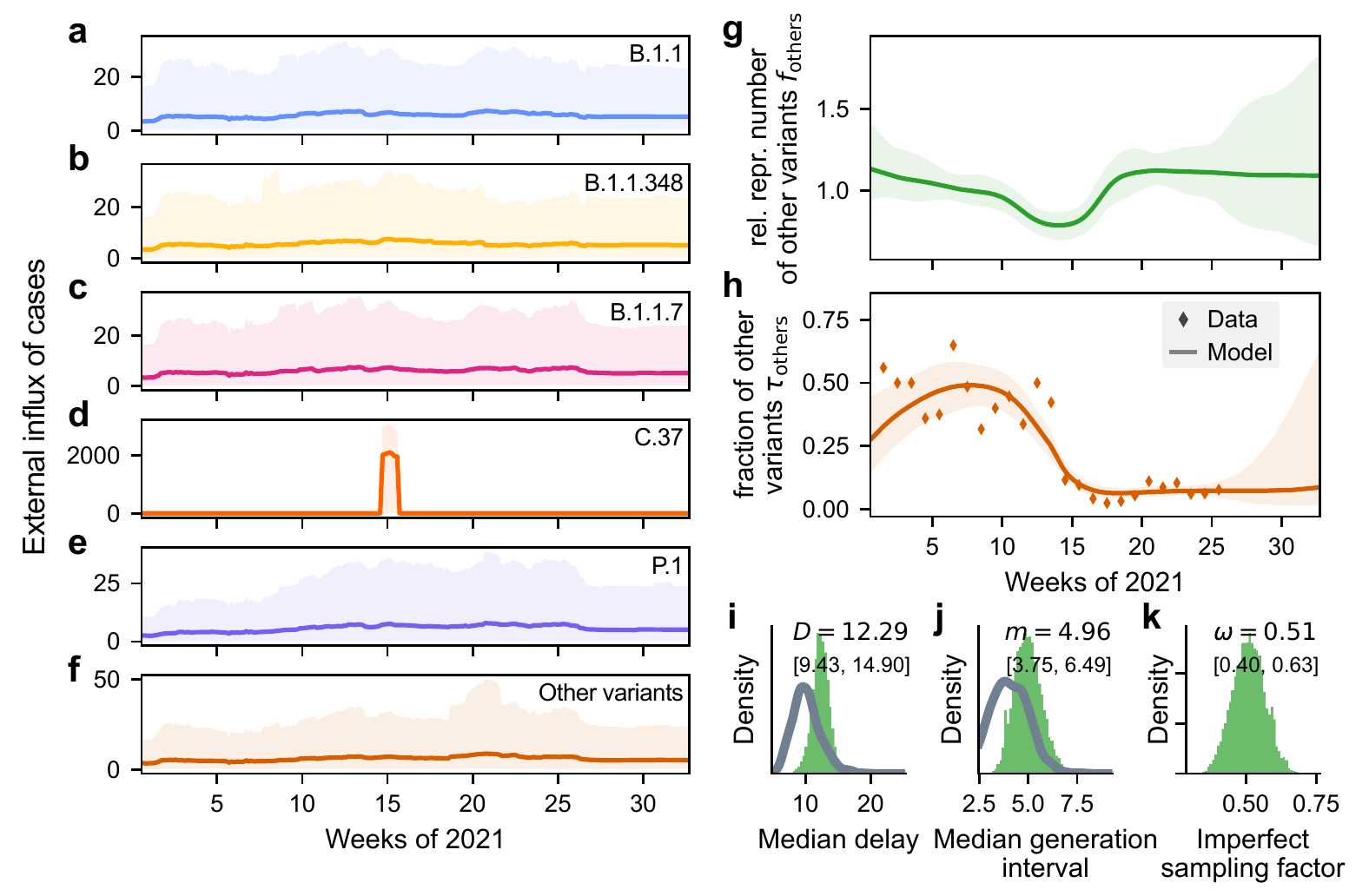}
    \caption{%
        \textbf{Posterior distributions of further parameters of the Bayesian model.} \textbf{a-f}: The external influx is low for most variants, following approximately our prior assumptions. An exception is Lambda, which eventually features a large influx when the measured fraction increased. Note however that the credible intervals of this influx are large, meaning that the model cannot decide whether the sudden increase Lambda cases is due to a large influx, or to a previous subsampling of Lambda cases (compare with \figref{fig:Figure_1} g). \textbf{g-h}: The relative reproduction number (compared to Alpha) of the modeled `other variants' spreading in Chile. These are the variants that are not separately modeled, and therefore are allowed to change their relative reproduction number over time. \textbf{i-k} Prior (gray) and posterior (green) distributions of some other parameters of the model.  
        }
    \label{fig:Figure_S2}
\end{figure}

\begin{table}[]
\centering
\caption{\textbf{Characteristic mutations in prevalent lineages.}}
\label{tab:Table_S1}
\begin{tabular}{lllllll}
\toprule
                & \multicolumn{6}{c}{\textbf{Genes}}                                                                                                                                                                                                                                                                                                                                                                                                                              \\ \midrule
\textbf{Lineage}         & \textbf{ORF1a}                                                                                                      & \textbf{ORF1b}                                                  & \textbf{S}                                                                                                                                    & \textbf{ORF3a} & \textbf{ORF8}                                                       & \textbf{N}                                                                    \\ \midrule
\textbf{B.1.1}           &                                                                                                            & P314L                                                  & D614G                                                                                                                                &       &                                                            & \begin{tabular}[c]{@{}l@{}}R203K\\ G204R\end{tabular}                \\ \midrule
\textbf{B.1.1.348}       & \begin{tabular}[c]{@{}l@{}}L1175F\\ V3718F\end{tabular}                                                    & P314L                                                  & \begin{tabular}[c]{@{}l@{}}D614G\\ G1167A\end{tabular}                                                                               & G174D &                                                            & \begin{tabular}[c]{@{}l@{}}S2Y\\ R203K\\ G204R\end{tabular}          \\ \midrule
\textbf{Alpha} & \begin{tabular}[c]{@{}l@{}}T1001I\\ A1708D\\ I2230T\\ del3675/3677\\ P314L\end{tabular}                    &                                                        & \begin{tabular}[c]{@{}l@{}}del69/70\\ del144/145\\ N501Y\\ A570D\\ D614G\\ P681H\\ T716I\\ S982A\\ D1118H\end{tabular}               &       & \begin{tabular}[c]{@{}l@{}}Q27*\\ R52I\\ Y73C\end{tabular} & \begin{tabular}[c]{@{}l@{}}D3L\\ R203K\\ G204R\\ S235F\end{tabular}  \\ \midrule
\textbf{Lambda}   & \begin{tabular}[c]{@{}l@{}}T1246I\\ P2287S\\ F2387V\\ L3201P\\ T3255I\\ G3278S\\ del3675/3677\end{tabular} & P314L                                                  & \begin{tabular}[c]{@{}l@{}}G75V\\ T76I\\ del247/253\\ L452Q\\ F490S\\ D614G\\ T859N\end{tabular}                                     &       &                                                            & \begin{tabular}[c]{@{}l@{}}P13L\\ R203K\\ G204R\\ G214C\end{tabular} \\ \midrule
\textbf{Gamma}     & \begin{tabular}[c]{@{}l@{}}S1188L\\ K1795Q\\ del3675/3677\end{tabular}                                     & \begin{tabular}[c]{@{}l@{}}P314L\\ E1264D\end{tabular} & \begin{tabular}[c]{@{}l@{}}L18F\\ T20N\\ P26S\\ D138Y\\ R190S\\ K417T\\ E484K\\ N501Y\\ D614G\\ H655Y\\ T1027I\\ V1176F\end{tabular} & S253P & E92K                                                       & \begin{tabular}[c]{@{}l@{}}P80R\\ R203K\\ G204R\end{tabular}         \\ \bottomrule
\end{tabular}%
\end{table}

\begin{table}[ht]
\centering
\caption{\textbf{Statistical assessment of mutation enrichment in the Spike gene.}}
\label{tab:Table_S2}
\begin{tabular}{lllll}
\toprule
Lineage & \makecell[c]{Levene \\ test} & \makecell[c]{Kolmogorov-Smirnov \\ test} & \makecell[c]{t-test with \\ different variances}& U-Test \\ \midrule
{Gamma}                         & 6.34e-19                                & 0.0                                                  & 0.0                                                         & 0.0                                 \\
{Lambda}                        & 0.0$^{*}$                                & 0.0                                                  & 0.0                                                         & 0.0$^{*}$                          \\
{Alpha}                     & 5.11e-06                                & 0.0$^{*}$                                            & 0.0$^{*}$                                                  & 0.0$^{*}$                           \\
{B.1.1}                       & 8.49e-05                                & 7.87e-05                                            & 0.0001                                                      & 0.0018                              \\
{B.1.1.348}                   & 0.0$^{*}$                                & 0.0$^{*}$                                            & 0.0$^{*}$                                                   & 0.0$^{*}$                           \\ \bottomrule
\multicolumn{5}{l}{$^{1}$\footnotesize{Values lower than $10^{-20}$ were considered as zero.}}
\end{tabular}
\end{table}

\begin{table}[h]
    \centering
    \caption{\textbf{Overview of model parameters.}}
    \begin{tabular}{cl}
        \toprule
        \textbf{Variable} & \textbf{Parameter} \\
        \midrule
        $R$                     & Effective Reproduction number \\
        $E$                     & New infectious\\
        $S$                     & Susceptible pool\\
        $g$                     & Generation interval\\
        $\Phi$                  & External influx\\
        $N$                     & Population size (19276715)\\
        $D$                     & Delay of case detection \\
        $M$                     & Reported (summed) cases in neighboring countries\\
        $d$                     & Length of change point\\
        $l$                     & Transient length of change point\\
        $\Gamma_c$              & Log-transformed reproduction number of each change point $c$\\
        $y$                     & Measured number of samples sequenced \\
        $n$                     & Total number of sequenced samples \\
        $\tau$                  & Fraction of variants in circulation \\
        $f$                     & Contribution of variant to spread\\
        $h_w$                   & Amplitude of weekend corrections\\
        $\chi_w$                & Phase shift of weekend correction\\
        Subscript $v$           & Denotes a distinct variant\\
        Subscript $t$           & Denotes discretized time\\
        Subscript $c$           & Denotes a change poit\\
    \end{tabular}
    \label{tab:parameters}
\end{table}

\begin{table}[h]
    \centering
    \caption{\textbf{List of priors.}}
    \begin{tabular}{cl}
        \toprule
        \textbf{Variable} & \textbf{Parameter} \\
        \midrule
        $E_{v,0}$               & $\text{HalfCauchy}(\sigma=100) \quad\forall v$\\
        $g_t$                   & $\text{LogNormal}(t; \mu=m,\sigma=0.4)$\\
        $m$                     & $\text{Normal}(\mu=4,\sigma=1)$\\
        $f_v$                   & $\text{LogNormal}(\mu=0,\sigma=1) \quad\forall v$\\
        $\Phi_{v,t}$            & $\text{HalfStudentT}_{\nu=4}(\sigma=0.0005) \quad\forall v,t$\\
        $\omega$                & $\text{Gamma}(\alpha=5,\beta=5)$\\
        $d_c$                   & $\text{Normal}(\mu=14c, \sigma=5) \quad\forall c$\\
        $l_c$                   & $\text{Normal}(\mu=20,\sigma=6) \quad\forall c$\\
        $\Gamma_c$              & $\text{Normal}(\mu=0,\sigma=0.2) + \Gamma_{c-1} \quad\forall c\neq0$\\
        $\Gamma_0$              & $\text{Normal}(\mu=1,\sigma=0.2)$\\
        $\kappa$                & $\text{HalfCauchy}(\sigma=10)$
    \end{tabular}
    \label{tab:priors}
\end{table}

\begin{table}[]
\centering
\caption{\textbf{Mutations becoming extinct and more predominant during vaccination roll-out in non-Spike proteins.}}
\label{tab:Table_S5}
\resizebox{\textwidth}{!}{%
\begin{tabular}{lll}
\hline
\textbf{Lineage} &
  \begin{tabular}[c]{@{}l@{}}\textbf{Synonymous mutations}\\ \textbf{becoming extinct}\end{tabular} &
  \begin{tabular}[c]{@{}l@{}}\textbf{Non-synonymous mutations}\\ \textbf{becoming more predominant}\end{tabular} \\ \hline
\textbf{B.1.1} &
  \begin{tabular}[c]{@{}l@{}}E\_V14V, NSP3\_F106F, NSP13\_L438L,\\ NSP4\_N244N\end{tabular} &
  \begin{tabular}[c]{@{}l@{}}NSP13\_E341D, NSP3\_A231V, NSP3\_A579V,\\ NSP3\_P1469S, NSP4\_L438P,  NSP4\_T492I,\\ NSP5\_G15S, NSP6\_S106-F108del, NSP8\_T141M,\\ N\_P80R, N\_S202T, N\_S235F,  ORF3a\_S253P,\\ ORF8\_E92K, ORF8\_Q27*\end{tabular} \\ \hline
\textbf{B.1.1.348} &
  \begin{tabular}[c]{@{}l@{}}NSP12\_N733N, NSP13\_L438L, NSP3\_F106F,\\ NSP4\_A416A, N\_R203R, ORF3a\_F43F\end{tabular} &
  \begin{tabular}[c]{@{}l@{}}NSP12\_I695T, NSP1\_L27F, NSP3\_A1215T,\\ NSP3\_K1386N, NSP3\_T678I, NSP9\_G38S,\end{tabular} \\ \hline
\textbf{Gamma} &
  \begin{tabular}[c]{@{}l@{}}NSP12\_D139D, NSP1\_D156D, NSP3\_D10D,\\ NSP3\_F106F, NSP3\_P1200P, NSP3\_V1298V,\\ NSP9\_Y31Y, N\_R203R\end{tabular} &
  None \\ \hline
\textbf{Alpha} &
  \begin{tabular}[c]{@{}l@{}}NSP12\_P411P, NSP2\_S36S, NSP3\_F106F,\\ NSP3\_F1089F\end{tabular} &
  NSP3\_F709L, NSP6\_L260F, N\_R203fs, \\ \hline
\textbf{Lambda} &
  None &
  None \\ \hline
\end{tabular}%
}
\end{table}


\begin{thebibliography}{10}

\bibitem{contreras2021risking}
Sebastian Contreras and Viola Priesemann.
\newblock Risking further {COVID-19} waves despite vaccination.
\newblock {\em The Lancet Infectious Diseases}, 2021.

\bibitem{bauer2021relaxing}
Simon Bauer, Sebastian Contreras, Jonas Dehning, Matthias Linden, Emil
  Iftekhar, Sebastian~B Mohr, {\'A}lvaro Olivera-Nappa, and Viola Priesemann.
\newblock Relaxing restrictions at the pace of vaccination increases freedom
  and guards against further covid-19 waves.
\newblock {\em arXiv preprint arXiv:2103.06228}, 2021.

\bibitem{viana2021controlling}
Joao Viana, Christiaan~H van Dorp, Ana Nunes, Manuel~C Gomes, Michiel van
  Boven, Mirjam~E Kretzschmar, Marc Veldhoen, and Ganna Rozhnova.
\newblock Controlling the pandemic during the {SARS-CoV-2} vaccination rollout:
  a modeling study.
\newblock {\em Nature communications}, 12(3674):1--15, 2021.

\bibitem{lavine2021immunological}
Jennie~S Lavine, Ottar~N Bjornstad, and Rustom Antia.
\newblock Immunological characteristics govern the transition of {COVID-19} to
  endemicity.
\newblock {\em Science}, 371(6530):741--745, 2021.

\bibitem{cobey2021concerns}
Sarah Cobey, Daniel~B Larremore, Yonatan~H Grad, and Marc Lipsitch.
\newblock Concerns about sars-cov-2 evolution should not hold back efforts to
  expand vaccination.
\newblock {\em Nature Reviews Immunology}, 21(5):330--335, 2021.

\bibitem{contreras2020low}
Sebastian Contreras, Jonas Dehning, Sebastian~B Mohr, F~Paul Spitzner, and
  Viola Priesemann.
\newblock Low case numbers enable long-term stable pandemic control without
  lockdowns.
\newblock {\em medRxiv}, 2020.

\bibitem{contreras2021challenges}
Sebastian Contreras, Jonas Dehning, Matthias Loidolt, Johannes Zierenberg,
  F~Paul Spitzner, Jorge~H Urrea-Quintero, Sebastian~B Mohr, Michael Wilczek,
  Michael Wibral, and Viola Priesemann.
\newblock The challenges of containing {SARS-CoV-2} via test-trace-and-isolate.
\newblock {\em Nature communications}, 12(1):1--13, 2021.

\bibitem{thompson2021incidence_and_escape}
Robin~N Thompson, Edward~M. Hill, and Julia~R. Gog.
\newblock Sars-cov-2 incidence and vaccine escape.
\newblock {\em Lancet Infectious Diseases}, 2021.

\bibitem{plante2021variant_gambit}
Jessica~A. Plante, Brooke~M. Mitchell, Kenneth~S. Plante, Kari Debbink,
  Scott~C. Weaver, and Vineet~D. Menachery.
\newblock The variant gambit: {COVID’s} next move.
\newblock {\em Cell Host \& Microbe}, 2021.

\bibitem{van2020risk}
Debra Van~Egeren, Alexander Novokhodko, Madison Stoddard, Uyen Tran, Bruce
  Zetter, Michael Rogers, Bradley~L Pentelute, Jonathan~M Carlson, Mark~S
  Hixon, Diane Joseph-McCarthy, et~al.
\newblock Risk of evolutionary escape from neutralizing antibodies targeting
  {SARS-CoV-2} spike protein.
\newblock {\em medRxiv}, 2020.

\bibitem{GISAID}
{Shu, Yuelong and McCauley, John}.
\newblock Gisaid: Global initiative on sharing all influenza data – from
  vision to reality.
\newblock {\em Eurosurveillance}, 22(13), 2017.

\bibitem{cyranoski_alarming_2021}
David Cyranoski.
\newblock Alarming {COVID} variants show vital role of genomic surveillance.
\newblock {\em Nature}, 589(7842):337--338, January 2021.
\newblock Bandiera\_abtest: a Cg\_type: News Number: 7842 Publisher: Nature
  Publishing Group.

\bibitem{malick2021genomic}
M~Shaheen~S Malick and Helen Fernandes.
\newblock The genomic landscape of sars-cov-2: Surveillance of variants of
  concern.
\newblock {\em Advances in Molecular Pathology}, 2021.

\bibitem{bartlow2021cooperative}
Andrew~W Bartlow, Earl~A Middlebrook, Alicia~T Romero, and Jeanne~M Fair.
\newblock How cooperative engagement programs strengthen sequencing
  capabilities for biosurveillance and outbreak response.
\newblock {\em Frontiers in Public Health}, 9:163, 2021.

\bibitem{helmy2016limited}
Mohamed Helmy, Mohamed Awad, and Kareem~A Mosa.
\newblock Limited resources of genome sequencing in developing countries:
  challenges and solutions.
\newblock {\em Applied \& translational genomics}, 9:15--19, 2016.

\bibitem{mena2021science}
Gonzalo~E. Mena, Pamela~P. Martinez, Ayesha~S. Mahmud, Pablo~A. Marquet,
  Caroline~O. Buckee, and Mauricio Santillana.
\newblock Socioeconomic status determines covid-19 incidence and related
  mortality in santiago, chile.
\newblock {\em Science}, 2021.

\bibitem{gozzi2021estimating}
Nicol{\`o} Gozzi, Michele Tizzoni, Matteo Chinazzi, Leo Ferres, Alessandro
  Vespignani, and Nicola Perra.
\newblock Estimating the effect of social inequalities on the mitigation of
  covid-19 across communities in santiago de chile.
\newblock {\em Nature communications}, 12(1):1--9, 2021.

\bibitem{bennett2021all}
Magdalena Bennett.
\newblock All things equal? heterogeneity in policy effectiveness against
  covid-19 spread in chile.
\newblock {\em World development}, 137:105208, 2021.

\bibitem{freire2021heterogeneous_spread_chile}
Danton Freire-Flores, Nyna Llanovarced-Kawles, Anamaria Sanchez-Daza, and
  Álvaro Olivera-Nappa.
\newblock On the heterogeneous spread of covid-19 in chile.
\newblock {\em Chaos Solitons \& Fractals}, 2021.

\bibitem{contreras2020multigroup}
Sebasti{\'{a}}n Contreras, H~Andr{\'{e}}s Villavicencio, David Medina-Ortiz,
  Juan~Pablo Biron-Lattes, and {\'{A}}lvaro Olivera-Nappa.
\newblock {A multi-group SEIRA model for the spread of {COVID-19} among
  heterogeneous populations}.
\newblock {\em Chaos, Solitons {\&} Fractals}, 136:109925, 2020.

\bibitem{castillo2020geographical}
Andr{\'e}s~E Castillo, B{\'a}rbara Parra, Paz Tapia, Jaime Lagos, Loredana
  Arata, Alejandra Acevedo, Winston Andrade, Gabriel Leal, Carolina Tambley,
  Patricia Bustos, et~al.
\newblock Geographical distribution of genetic variants and lineages of
  sars-cov-2 in chile.
\newblock {\em Frontiers in public health}, 8:525, 2020.

\bibitem{shepherd2021vaccination_in_Chile}
Alison Shepherd.
\newblock Covid-19: Chile joins top five countries in world vaccination league.
\newblock {\em BMJ}, 2021.

\bibitem{aguilera2021}
Ximena Aguilera, Adrian~P. Mundt, Rafael Araos, and Thomas Weitzel.
\newblock The story behind chile's rapid rollout of covid-19 vaccination.
\newblock {\em Travel Medicine and Infectious Disease}, 2021.

\bibitem{owidcoronavirus}
Esteban Ortiz-Ospina Max~Roser, Hannah~Ritchie and Joe Hasell.
\newblock Coronavirus pandemic (covid-19).
\newblock {\em Our World in Data}, 2020.
\newblock \url{https://ourworldindata.org/coronavirus},
  (\href{https://ourworldindata.org/coronavirus-data-explorer?zoomToSelection=true&time=2020-05-15..2020-09-15&country=DEU~CZE~ITA~AUT~BEL~DNK~FIN~FRA~NLD~POL~ESP~CHE&region=SouthAmerica&casesMetric=true&interval=smoothed&perCapita=true&smoothing=7&pickerMetric=new_cases&pickerSort=desc}{Europe},
  \href{https://ourworldindata.org/coronavirus-data-explorer?zoomToSelection=true&time=2020-05-15..2020-09-15&country=USA~ARG~BRA~CHL~COL~PER&region=SouthAmerica&casesMetric=true&interval=smoothed&perCapita=true&smoothing=7&pickerMetric=new_cases&pickerSort=desc}{America},
  and
  \href{https://ourworldindata.org/coronavirus-data-explorer?zoomToSelection=true&time=2020-05-15..2020-09-15&country=CHN~JPN~MYS~KOR~AUS~NZL&region=SouthAmerica&casesMetric=true&interval=smoothed&perCapita=true&smoothing=7&pickerMetric=new_cases&pickerSort=desc}{Oceania
  and Asia}).

\bibitem{asahi2021lockdowns_impacted_chile_economics}
Kenzo Asahi, Eduardo~A. Undurraga, Rodrigo Valdés, and Rodrigo Wagner.
\newblock The effect of covid-19 on the economy: Evidence from an early adopter
  of localized lockdowns.
\newblock {\em Journal of Global Health}, 2021.

\bibitem{contreras2020statistically}
Sebasti{\'{a}}n Contreras, Juan~Pablo Biron-Lattes, H~Andr{\'{e}}s
  Villavicencio, David Medina-Ortiz, Nyna Llanovarced-Kawles, and {\'{A}}lvaro
  Olivera-Nappa.
\newblock {Statistically-based methodology for revealing real contagion trends
  and correcting delay-induced errors in the assessment of {COVID-19}
  pandemic}.
\newblock {\em Chaos, Solitons {\&} Fractals}, 139:110087, 2020.

\bibitem{MINSAL2021trazabilidad}
Ministerio de Salud de Chile~(MINSAL) Department~of Epidemiology.
\newblock {Tech Report: National strategy for test-trace-and-isolate
  ({COVID-19}), 3--9, July, 2021 (Estrategia Nacional de Testeo, Trazabilidad y
  Aislamiento {COVID-19}, SEMANA DEL 3 - 9 DE JULIO, 2021)}.
\newblock
  \url{https://www.minsal.cl/wp-content/uploads/2021/07/Indicadores-de-Testeo-y-Trazabilidad-13072021.pdf}.

\bibitem{minsal_2021}
MINSAL MINSAL.
\newblock Vacunas contra sars- cov-2 utilizadas en chile mantienen altos
  niveles de efectividad para evitar hospitalización, ingreso a uci y muerte,
  Aug 2021.

\bibitem{jara2021effectiveness}
Alejandro Jara, Eduardo~A Undurraga, Cecilia Gonz{\'a}lez, Fabio Paredes,
  Tom{\'a}s Fontecilla, Gonzalo Jara, Alejandra Pizarro, Johanna Acevedo,
  Katherine Leo, Francisco Leon, et~al.
\newblock Effectiveness of an inactivated sars-cov-2 vaccine in chile.
\newblock {\em New England Journal of Medicine}, 2021.

\bibitem{gonzalez2021mutations_in_magallanes}
Jorge González-Puelma, Jacqueline Aldridge, Marco~Montes de~Oca, Mónica
  Pinto, Roberto Uribe-Paredes, Jose Fernandez-Goycoolea, Diego
  Alvarez-Saravia, Hermy Álvarez, Gonzalo Encina, Thomas Weitzel, Thomas
  Weitzel, Rodrigo Muñoz, Rodrigo Muñoz, Álvaro Olivera-Nappa, Sergio
  Pantano, Sergio Pantano, and Marcelo~A. Navarrete.
\newblock Mutation in a sars-cov-2 haplotype from sub-antarctic chile reveals
  new insights into the spike’s dynamics.
\newblock {\em Viruses}, 2021.

\bibitem{acevedo2021lambda_infectivity_chile}
M{\'o}nica~L Acevedo, Luis Alonso-Palomares, Andr{\'e}s Bustamante, Aldo
  Gaggero, Fabio Paredes, Claudia~P Cort{\'e}s, Fernando
  Valiente-Echeverr{\'\i}a, and Ricardo Soto-Rifo.
\newblock Infectivity and immune escape of the new sars-cov-2 variant of
  interest lambda.
\newblock {\em medRxiv}, 2021.

\bibitem{romero2021lambda_in_peru}
Pedro~E Romero, Alejandra D{\'a}vila-Barclay, Guillermo Salvatierra, Luis
  Gonz{\'a}lez, Diego Cuicapuza, Luis Solis, Pool Marcos-Carbajal, Janet
  Huancachoque, Lenin Maturrano, and Pablo Tsukayama.
\newblock The emergence of sars-cov-2 variant lambda (c. 37) in south america.
\newblock {\em medRxiv}, 2021.

\bibitem{mora2021emergencia}
Eduardo~Lopez Mora, Jorge Espinoza, Jeannette Dabanch, and Rodrigo Cruz.
\newblock Emergencia de variante delta-b. 1.617. 2. su impacto potencial en la
  evoluci{\'o}n de la pandemia por sars-cov-2.
\newblock {\em Bolet{\'\i}n Micol{\'o}gico}, 36(1), 2021.

\bibitem{vargas_2021}
Pilar Vargas.
\newblock Comunicado de sochinf sobre variante delta en chile, Jul 2021.

\bibitem{shepard_viral_2016}
Samuel~S. Shepard, Sarah Meno, Justin Bahl, Malania~M. Wilson, John Barnes, and
  Elizabeth Neuhaus.
\newblock Viral deep sequencing needs an adaptive approach: {IRMA}, the
  iterative refinement meta-assembler.
\newblock {\em BMC Genomics}, 17(1):708, September 2016.

\bibitem{katoh_mafft_2002}
Kazutaka Katoh, Kazuharu Misawa, Kei‐ichi Kuma, and Takashi Miyata.
\newblock {MAFFT}: a novel method for rapid multiple sequence alignment based
  on fast {Fourier} transform.
\newblock {\em Nucleic Acids Research}, 30(14):3059--3066, July 2002.

\bibitem{rambaut_dynamic_2020}
Andrew Rambaut, Edward~C. Holmes, Áine O’Toole, Verity Hill, John~T.
  McCrone, Christopher Ruis, Louis du~Plessis, and Oliver~G. Pybus.
\newblock A dynamic nomenclature proposal for {SARS}-{CoV}-2 lineages to assist
  genomic epidemiology.
\newblock {\em Nature Microbiology}, 5(11):1403--1407, November 2020.
\newblock Bandiera\_abtest: a Cg\_type: Nature Research Journals Number: 11
  Primary\_atype: Research Publisher: Nature Publishing Group Subject\_term:
  Classification and taxonomy;Phylogenetics;Phylogeny;SARS-CoV-2;Viral
  evolution Subject\_term\_id:
  classification-and-taxonomy;phylogenetics;phylogeny;sars-cov-2;viral-evolution.

\bibitem{fraser_estimating_2007}
Christophe Fraser.
\newblock Estimating {Individual} and {Household} {Reproduction} {Numbers} in
  an {Emerging} {Epidemic}.
\newblock {\em PLoS ONE}, 2(8), August 2007.

\bibitem{Flaxman2020estimating}
Seth Flaxman, Swapnil Mishra, Axel Gandy, H~Juliette~T Unwin, Thomas~A Mellan,
  Helen Coupland, Charles Whittaker, Harrison Zhu, Tresnia Berah, Jeffrey~W
  Eaton, and Others.
\newblock {Estimating the effects of non-pharmaceutical interventions on
  {COVID-19} in Europe}.
\newblock {\em Nature}, pages 1--8, 2020.

\bibitem{brauner_inferring_2020}
Jan~M. Brauner, Sören Mindermann, Mrinank Sharma, David Johnston, John
  Salvatier, Tomáš Gavenčiak, Anna~B. Stephenson, Gavin Leech, George
  Altman, Vladimir Mikulik, Alexander~John Norman, Joshua~Teperowski Monrad,
  Tamay Besiroglu, Hong Ge, Meghan~A. Hartwick, Yee~Whye Teh, Leonid
  Chindelevitch, Yarin Gal, and Jan Kulveit.
\newblock Inferring the effectiveness of government interventions against
  {COVID}-19.
\newblock {\em Science}, 2020.

\bibitem{dehning2020inferring}
Jonas Dehning, Johannes Zierenberg, F~Paul Spitzner, Michael Wibral,
  Joao~Pinheiro Neto, Michael Wilczek, and Viola Priesemann.
\newblock {Inferring change points in the spread of {COVID-19} reveals the
  effectiveness of interventions}.
\newblock {\em Science}, 2020.

\bibitem{davies_estimated_2021}
Nicholas~G. Davies, Sam Abbott, Rosanna~C. Barnard, Christopher~I. Jarvis,
  Adam~J. Kucharski, James~D. Munday, Carl A.~B. Pearson, Timothy~W. Russell,
  Damien~C. Tully, Alex~D. Washburne, Tom Wenseleers, Amy Gimma, William
  Waites, Kerry L.~M. Wong, Kevin~van Zandvoort, Justin~D. Silverman, CMMID
  COVID-19~Working Group1‡, COVID-19 Genomics UK (COG-UK) Consortium‡,
  Karla Diaz-Ordaz, Ruth Keogh, Rosalind~M. Eggo, Sebastian Funk, Mark Jit,
  Katherine~E. Atkins, and W.~John Edmunds.
\newblock Estimated transmissibility and impact of {SARS}-{CoV}-2 lineage
  {B}.1.1.7 in {England}.
\newblock {\em Science}, March 2021.

\bibitem{volz_bayesian_2018}
Erik~M. Volz and Igor Siveroni.
\newblock Bayesian phylodynamic inference with complex models.
\newblock {\em PLOS Computational Biology}, 14(11):e1006546, November 2018.

\bibitem{bouckaert_beast_2019}
Remco Bouckaert, Timothy~G. Vaughan, Joëlle Barido-Sottani, Sebastián
  Duchêne, Mathieu Fourment, Alexandra Gavryushkina, Joseph Heled, Graham
  Jones, Denise Kühnert, Nicola~De Maio, Michael Matschiner, Fábio~K. Mendes,
  Nicola~F. Müller, Huw~A. Ogilvie, Louis~du Plessis, Alex Popinga, Andrew
  Rambaut, David Rasmussen, Igor Siveroni, Marc~A. Suchard, Chieh-Hsi Wu, Dong
  Xie, Chi Zhang, Tanja Stadler, and Alexei~J. Drummond.
\newblock {BEAST} 2.5: {An} advanced software platform for {Bayesian}
  evolutionary analysis.
\newblock {\em PLOS Computational Biology}, 15(4):e1006650, April 2019.

\bibitem{faria_genomics_2021}
Nuno~R. Faria, Thomas~A. Mellan, Charles Whittaker, Ingra~M. Claro, Darlan
  da~S. Candido, Swapnil Mishra, Myuki A.~E. Crispim, Flavia C.~S. Sales, Iwona
  Hawryluk, John~T. McCrone, Ruben J.~G. Hulswit, Lucas A.~M. Franco,
  Mariana~S. Ramundo, Jaqueline G.~de Jesus, Pamela~S. Andrade, Thais~M.
  Coletti, Giulia~M. Ferreira, Camila A.~M. Silva, Erika~R. Manuli, Rafael
  H.~M. Pereira, Pedro~S. Peixoto, Moritz U.~G. Kraemer, Nelson Gaburo, Cecilia
  da~C. Camilo, Henrique Hoeltgebaum, William~M. Souza, Esmenia~C. Rocha,
  Leandro M.~de Souza, Mariana C.~de Pinho, Leonardo J.~T. Araujo, Frederico
  S.~V. Malta, Aline B.~de Lima, Joice do~P. Silva, Danielle A.~G. Zauli,
  Alessandro C. de~S. Ferreira, Ricardo~P. Schnekenberg, Daniel~J. Laydon,
  Patrick G.~T. Walker, Hannah~M. Schlüter, Ana L. P.~dos Santos, Maria~S.
  Vidal, Valentina S.~Del Caro, Rosinaldo M.~F. Filho, Helem M.~dos Santos,
  Renato~S. Aguiar, José~L. Proença-Modena, Bruce Nelson, James~A. Hay,
  Mélodie Monod, Xenia Miscouridou, Helen Coupland, Raphael Sonabend, Michaela
  Vollmer, Axel Gandy, Carlos~A. Prete, Vitor~H. Nascimento, Marc~A. Suchard,
  Thomas~A. Bowden, Sergei L.~K. Pond, Chieh-Hsi Wu, Oliver Ratmann, Neil~M.
  Ferguson, Christopher Dye, Nick~J. Loman, Philippe Lemey, Andrew Rambaut,
  Nelson~A. Fraiji, Maria do P. S.~S. Carvalho, Oliver~G. Pybus, Seth Flaxman,
  Samir Bhatt, and Ester~C. Sabino.
\newblock Genomics and epidemiology of the {P}.1 {SARS}-{CoV}-2 lineage in
  {Manaus}, {Brazil}.
\newblock {\em Science}, 372(6544):815--821, May 2021.
\newblock Publisher: American Association for the Advancement of Science
  Section: Research Article.

\bibitem{Cov_lineages_B117}
Áine O'Toole, Verity Hill, and GISAID.
\newblock {COV-lineages: B.1.1.7}.
\newblock \url{https://cov-lineages.org/global_report_B.1.1.7}.

\bibitem{teng_systemic_2020}
Shaolei Teng, Adebiyi Sobitan, Raina Rhoades, Dongxiao Liu, and Qiyi Tang.
\newblock Systemic effects of missense mutations on {SARS}-{CoV}-2 spike
  glycoprotein stability and receptor-binding affinity.
\newblock {\em Briefings in Bioinformatics}, 22(2):1239--1253, October 2020.

\bibitem{Cov_lineages_B11}
{Lineage Mutation Tracker from Outbreak.info: B.1.1}.
\newblock \url{https://outbreak.info/situation-reports?pango=B.1.1}.

\bibitem{wang_characterizing_2020}
Rui Wang, Jiahui Chen, Kaifu Gao, Yuta Hozumi, Changchuan Yin, and Guo-Wei Wei.
\newblock Characterizing {SARS}-{CoV}-2 mutations in the {United} {States}.
\newblock {\em Research Square}, pages rs.3.rs--49671, August 2020.

\bibitem{mohammadi_novel_2021}
Elmira Mohammadi, Fatemeh Shafiee, Kiana Shahzamani, Mohammad~Mehdi Ranjbar,
  Abbas Alibakhshi, Shahrzad Ahangarzadeh, Leila Beikmohammadi, Laleh Shariati,
  Soodeh Hooshmandi, Behrooz Ataei, and Shaghayegh~Haghjooy Javanmard.
\newblock Novel and emerging mutations of {SARS}-{CoV}-2: {Biomedical}
  implications.
\newblock {\em Biomedicine \& Pharmacotherapy}, 139:111599, July 2021.

\bibitem{volz_assessing_2021}
Erik Volz, Swapnil Mishra, Meera Chand, Jeffrey~C. Barrett, Robert Johnson,
  Lily Geidelberg, Wes~R. Hinsley, Daniel~J. Laydon, Gavin Dabrera, Áine
  O’Toole, Robert Amato, Manon Ragonnet-Cronin, Ian Harrison, Ben Jackson,
  Cristina~V. Ariani, Olivia Boyd, Nicholas~J. Loman, John~T. McCrone, Sónia
  Gonçalves, David Jorgensen, Richard Myers, Verity Hill, David~K. Jackson,
  Katy Gaythorpe, Natalie Groves, John Sillitoe, Dominic~P. Kwiatkowski, Seth
  Flaxman, Oliver Ratmann, Samir Bhatt, Susan Hopkins, Axel Gandy, Andrew
  Rambaut, and Neil~M. Ferguson.
\newblock Assessing transmissibility of {SARS}-{CoV}-2 lineage {B}.1.1.7 in
  {England}.
\newblock {\em Nature}, 593(7858):266--269, May 2021.
\newblock Bandiera\_abtest: a Cg\_type: Nature Research Journals Number: 7858
  Primary\_atype: Research Publisher: Nature Publishing Group Subject\_term:
  Population genetics;SARS-CoV-2;Viral infection Subject\_term\_id:
  population-genetics;sars-cov-2;viral-infection.

\bibitem{frampton_genomic_2021}
Dan Frampton, Tommy Rampling, Aidan Cross, Heather Bailey, Judith Heaney,
  Matthew Byott, Rebecca Scott, Rebecca Sconza, Joseph Price, Marios
  Margaritis, Malin Bergstrom, Moira~J. Spyer, Patricia~B. Miralhes, Paul
  Grant, Stuart Kirk, Chris Valerio, Zaheer Mangera, Thaventhran Prabhahar,
  Jeronimo Moreno-Cuesta, Nish Arulkumaran, Mervyn Singer, Gee~Yen Shin, Emilie
  Sanchez, Stavroula~M. Paraskevopoulou, Deenan Pillay, Rachel~A. McKendry,
  Mariyam Mirfenderesky, Catherine~F. Houlihan, and Eleni Nastouli.
\newblock Genomic characteristics and clinical effect of the emergent
  {SARS}-{CoV}-2 {B}.1.1.7 lineage in {London}, {UK}: a whole-genome sequencing
  and hospital-based cohort study.
\newblock {\em The Lancet Infectious Diseases}, 0(0), April 2021.
\newblock Publisher: Elsevier.

\bibitem{chakraborty_evolutionary_2021}
Sandipan Chakraborty.
\newblock Evolutionary and structural analysis elucidates mutations on
  {SARS}-{CoV2} spike protein with altered human {ACE2} binding affinity.
\newblock {\em Biochemical and Biophysical Research Communications},
  538:97--103, January 2021.

\bibitem{wink_first_2021}
Priscila~Lamb Wink, Fabiana Caroline~Zempulski Volpato, Francielle~Liz
  Monteiro, Julia~Biz Willig, Alexandre~Prehn Zavascki, Afonso~Luís Barth, and
  Andreza~Francisco Martins.
\newblock First identification of {SARS}-{CoV}-2 {Lambda} ({C}.37) variant in
  {Southern} {Brazil}.
\newblock {\em medRxiv}, page 2021.06.21.21259241, June 2021.
\newblock Publisher: Cold Spring Harbor Laboratory Press.

\bibitem{noauthor_outbreakinfo_nodate}
outbreak.info.

\bibitem{naveca_covid-19_2021}
Felipe~Gomes Naveca, Valdinete Nascimento, Victor~Costa de~Souza, André
  de~Lima Corado, Fernanda Nascimento, George Silva, Ágatha Costa, Débora
  Duarte, Karina Pessoa, Matilde Mejía, Maria~Júlia Brandão, Michele Jesus,
  Luciana Gonçalves, Cristiano~Fernandes da~Costa, Vanderson Sampaio, Daniel
  Barros, Marineide Silva, Tirza Mattos, Gemilson Pontes, Ligia Abdalla,
  João~Hugo Santos, Ighor Arantes, Filipe~Zimmer Dezordi, Marilda~Mendonça
  Siqueira, Gabriel~Luz Wallau, Paola~Cristina Resende, Edson Delatorre, Tiago
  Gräf, and Gonzalo Bello.
\newblock {COVID}-19 in {Amazonas}, {Brazil}, was driven by the persistence of
  endemic lineages and {P}.1 emergence.
\newblock {\em Nature Medicine}, pages 1--9, May 2021.
\newblock Bandiera\_abtest: a Cg\_type: Nature Research Journals
  Primary\_atype: Research Publisher: Nature Publishing Group Subject\_term:
  SARS-CoV-2;Virology Subject\_term\_id: sars-cov-2;virology.

\bibitem{nagy_different_2021}
Ádám Nagy, Sándor Pongor, and Balázs Győrffy.
\newblock Different mutations in {SARS}-{CoV}-2 associate with severe and mild
  outcome.
\newblock {\em International Journal of Antimicrobial Agents}, 57(2):106272,
  February 2021.

\bibitem{Campbell2021Increased}
Finlay Campbell, Brett Archer, Henry Laurenson-Schafer, Yuka Jinnai, Franck
  Konings, Neale Batra, Boris Pavlin, Katelijn Vandemaele, Maria~D
  Van~Kerkhove, Thibaut Jombart, Oliver Morgan, and Olivier le~Polain~de
  Waroux.
\newblock Increased transmissibility and global spread of sars-cov-2 variants
  of concern as at june 2021.
\newblock {\em Eurosurveillance}, 26(24), 2021.

\bibitem{rochman_ongoing_2021}
Nash~D. Rochman, Yuri~I. Wolf, Guilhem Faure, Pascal Mutz, Feng Zhang, and
  Eugene~V. Koonin.
\newblock Ongoing global and regional adaptive evolution of {SARS}-{CoV}-2.
\newblock {\em Proceedings of the National Academy of Sciences}, 118(29), July
  2021.
\newblock Publisher: National Academy of Sciences Section: Biological Sciences.

\bibitem{harvey_sars-cov-2_2021}
William~T. Harvey, Alessandro~M. Carabelli, Ben Jackson, Ravindra~K. Gupta,
  Emma~C. Thomson, Ewan~M. Harrison, Catherine Ludden, Richard Reeve, Andrew
  Rambaut, Sharon~J. Peacock, and David~L. Robertson.
\newblock {SARS}-{CoV}-2 variants, spike mutations and immune escape.
\newblock {\em Nature Reviews Microbiology}, 19(7):409--424, July 2021.
\newblock Bandiera\_abtest: a Cg\_type: Nature Research Journals Number: 7
  Primary\_atype: Reviews Publisher: Nature Publishing Group Subject\_term:
  Protein analysis;SARS-CoV-2;Vaccines;Viral evolution;Viral infection
  Subject\_term\_id:
  protein-analysis;sars-cov-2;vaccines;viral-evolution;viral-infection.

\bibitem{acevedo_infectivity_2021}
Mónica~L. Acevedo, Luis Alonso-Palomares, Andrés Bustamante, Aldo Gaggero,
  Fabio Paredes, Claudia~P. Cortés, Fernando Valiente-Echeverría, and Ricardo
  Soto-Rifo.
\newblock Infectivity and immune escape of the new {SARS}-{CoV}-2 variant of
  interest {Lambda}.
\newblock {\em medRxiv}, page 2021.06.28.21259673, July 2021.
\newblock Publisher: Cold Spring Harbor Laboratory Press.

\bibitem{JHU}
E.~Dong, H.~Du, and L.~Gardner.
\newblock An interactive web-based dashboard to track covid-19 in real time.
\newblock {\em The Lancet. Infectious Diseases}, 20:533 -- 534, 2020.

\end{thebibliography}
\end{document}